\documentclass[openacc]{rstransa}

\usepackage{amsmath}
\usepackage{amssymb}
\usepackage{graphicx}

\usepackage[numbers,sort&compress]{natbib}

\newcommand*{\doi}[1]{\href{http://dx.doi.org/#1}{doi: #1}}

\usepackage{hyperref}
\usepackage{enumitem}

\usepackage{tcolorbox}
\definecolor{abstract-color}{cmyk}{0.04, 0.04, 0.12, 0.08}

\begin{document}

\title{High-frequency oscillations in small chromospheric bright features observed with ALMA}

\author{
J.~C.~Guevara G\'omez$^{1,2}$, S.~Jafarzadeh$^{1,2}$, S.~Wedemeyer$^{1,2}$, M.~Szydlarski$^{1,2}$, M.~Stangalini$^{3}$, B. Fleck$^{4}$, and P.~H.~Keys$^{5}$}

\address{$^{1}$Rosseland Centre for Solar Physics, University of Oslo, P.O. Box 1029 Blindern, NO-0315 Oslo, Norway\\
$^{2}$Institute of Theoretical Astrophysics, University of Oslo, P.O. Box 1029 Blindern, NO-0315 Oslo, Norway\\
$^{3}$ASI Italian Space Agency, Via del Politecnico snc, I-00133 Rome, Italy\\
$^{4}$ESA Science and Operations Department, c/o NASA Goddard Space Flight Center, Greenbelt, MD 20771, USA\\
$^{5}$Astrophysics Research Centre, School of Mathematics and Physics, Queen’s University Belfast, Belfast, BT7 1NN, UK
}

\subject{astrophysics, observational astronomy, solar system, wave motion}


\corres{J.C.~Guevara~G\'omez\\
\email{j.c.g.gomez@astro.uio.no}}



\maketitle

\begin{tcolorbox}[sharp corners, width=\textwidth,colback=abstract-color,colframe=abstract-color,boxsep=5pt,left=0pt,right=0pt,top=0pt,bottom=0pt]
We report detection of oscillations in brightness temperature, size, and horizontal velocity of three small bright features in the chromosphere of a plage/enhanced-network region. The observations, which were taken with high temporal resolution (i.e., 2-sec cadence) with the Atacama Large Millimeter/submillimeter Array (ALMA) in Band~3 (centred at 3 mm; 100 GHz), exhibit three small-scale features with oscillatory behaviour with different, but overlapping, distributions of period on the order of, on average, $90 \pm 22$~s, $110 \pm 12$~s and $66 \pm 23$~s, respectively. We find anti-correlations between perturbations in brightness temperature and size of the three features, which suggest the presence of fast sausage-mode waves in these small structures. In addition, the detection of transverse oscillations (although with a larger uncertainty) may suggest as well the presence of Alfv\'enic oscillations which are likely representative of kink waves. This work demonstrates the diagnostic potential of high-cadence observations with ALMA for detecting high-frequency magnetohydrodynamic waves in the solar chromosphere. Such waves can potentially channel a vast amount of energy into the outer atmosphere of the Sun.
\end{tcolorbox}

\section{Introduction}

The solar atmosphere consists of a large variety of magnetic structures capable of maintaining 
different types of magnetohydrodynamic (MHD) waves \cite{2005SSRv..121..115N,2009SSRv..149..299V,2009SSRv..149...65D,2016GMS...216..449J}. While these could be in part responsible for the plasma heating in the upper solar atmosphere (by means of energy deposition in those layers), the exact physical mechanisms of which still remain unclear (see \cite{2015SSRv..190..103J,2019FrASS...6...48A} and references therein for a wider picture). In particular, direct detection of MHD-wave-energy release in the solar chromosphere has been rare, thus, their detection in structures with various spatial scales and in a wide range of frequencies are of vital importance.

Notably, compressible and incompressible MHD-wave modes can be excited in magnetic flux tubes, which act as waveguides, by reoccurring perturbations at their footpoints at photospheric heights \cite{2016GMS...216..431V}.
Signatures of both standing and propagating incompressible (transverse) kink waves, as oscillatory pattern of velocities perpendicular to the assumed waveguides, have been observed in various structures in the solar photosphere and chromosphere such as small-scale bright points, spicules, fibrils, and mottles  (e.g., \cite{2007Sci...318.1574D,2007SoPh..246...65L,2012ApJ...750...51K,2013A&A...554A.115S, 2014A&A...569A.102S,2015A&A...577A..17S,2017ApJ...840...19S,2017ApJS..229....9J,2017ApJS..229...10J}). These have been reported to have periods on the order of 30-350~s and velocity amplitudes of about 1-29~km\,s$^{-1}$ (see table 3 in \cite{2015SSRv..190..103J} for a detailed summary of the observed properties of kink waves).
In the case of the compressible sausage modes, the observed signature is a periodic fluctuation of the waveguide's cross-section accompanied by a corresponding out-of-phase oscillation of the intensity \cite{1983SoPh...88..179E,2009A&A...494..295E}. \citet{2011ApJ...729L..18M} reported evidence of the sausage mode in a magnetic pore with a clear anti-phase relation of the width of the magnetic structure (waveguide) and the intensity observed at 4170\,\AA~by the Rapid Oscillations in the Solar Atmosphere (ROSA; \cite{2010SoPh..261..363J}) instrument  at Dunn Solar Telescope. Furthermore, \citet{2017ApJS..229....7G} reported the anti-phase behaviour in Slender Ca~{\sc ii}~H Fibrils observed at 3969\,\AA~with the SuFI instrument onboard the {\sc Sunrise} balloon-borne solar observatory \cite{2010ApJ...723L.127S,2017ApJS..229....2S}.

Both kink- and sausage-mode waves have been observed in large magnetic structures such as sunspots and pores, as well as, in small-scale elongated structures. Their observations in small magnetic features, particularly in the mid-to-high solar chromosphere, have not been so common. Such waves, if propagating, can significantly contribute to the heating of the outer solar atmosphere where they can be dissipated. 
Thus, their detection in the chromosphere is of high interest for understanding the energy budget in this atmospheric region and beyond.

The Atacama Large Millimeter/submillimeter Array (ALMA; \cite{2009IEEEP..97.1463W}) started regular observations of the Sun in late 2016, with a capability of providing high-cadence observations of the solar chromosphere at millimeter wavelengths. This is obviously a crucial factor for observing high-frequency (short period) waves. These will, however, be limited to spatial scales detectable by currently provided antennas configurations of ALMA, since the spatial resolution is also important in detecting high-frequency oscillations \cite{2007ASPC..368...93W}. Additionally, ALMA is a powerful diagnostic tool for the solar chromosphere, since the radiation is formed under the condition of local thermodynamic equilibrium (LTE) at these wavelengths. Hence, the observable brightness temperature constitutes a direct proxy of the local gas temperature of the plasma \cite{2002AN....323..271B,2016Msngr.163...15W,2016SSRv..200....1W,2017A&A...601A..43L, 2017SoPh..292...87S}. Thus, temperature perturbations due to, e.g., propagating waves, can be directly examined.

Recent studies have shown the potential of ALMA to examine distinct phenomena both in quiet and active regions \cite{2017ApJ...841L...5S,2018ApJ...863...96Y,2018A&A...619L...6N,2019A&A...622A.150J,2019ApJ...877L..26L,2020A&A...634A..86P,2020A&A...634A..56D,2020ApJ...888L..28S,2020A&A...635A..71W,2020A&A...638A..62N}. Particularly, \citet{Eklund2020} found signatures of shock-wave events in Band 3 (2.8-3.3\,mm) observations from December 2016, coming to the conclusion that there are numerous small-scale dynamic structures with lifetimes of 43-360 s present in the ALMA field of view (FOV), with excess temperatures of more than 400\,K and a correlation between their occurrence and the magnetic field strength. Characterisation of such events at millimetre wavelengths have also been studied in detail from numerical simulations \cite{Eklund2020RS}. \citet{2020A&A...638A..62N} presented a survey of transient brightenings with a mean lifetime of 51.1 s in ALMA Band 3 data observed in March 2017, which show light curves such that their origin is strongly suggested to be thermal, meaning that the observed brightness temperature variations are indeed caused by variations of the gas temperature and not other effects such as optical depth variations. Lately, \citet{2020arXiv200512717C} reported the detection of an on-disk chromospheric spicule in a plage region observed simultaneously by IRIS and ALMA Band 6 (1.1-1.4 \,mm) in April 2017. Therefore, ALMA observations have been demonstrated to possess sufficient quality and temporal resolution to search for various long-lived and transient phenomena (within its current spatial-resolution limits), and likewise, to study MHD-wave signatures in the solar chromosphere. An overall study of temperature oscillations in the solar chromosphere with ALMA (using ten different datasets in both Band~3 and Band~6) has shown the dependency of the oscillation properties on the magnetic configurations in the mid-to-high chromosphere \cite{Jafarzadeh2020}. Such oscillations (in Band 6) have shown not to be correlated with those observed in the low chromosphere \cite{Narang2020}.

In this paper we present evidence of MHD-wave signatures in three small bright features traced in time within the FOV of an ALMA Band 3 data set. In Section~\ref{sec:obs}, the data is described. The method of selection of the three bright features as well as the tracking algorithm are presented in Section~\ref{sec:method}. In Section~\ref{sec:analysis}, the properties of the features and the wave analysis are detailed. Lastly, our interpretation of the observed oscillations as well our concluding remarks are presented in Section~\ref{sec:dis_conc}.

\section{Observations\label{sec:obs}}

The ALMA Band 3 (2.8-3.3\,mm) observations used here were carried out on 2017 April 22nd between 17:20 and 17:54 UTC as part of program 2016.1.00050.S. The setup of the receiver bands consists of four sub-bands centred on 93\,GHz (SB1), 95\,GHz (SB2), 105\,GHz (SB3), and 107\,GHz (SB4), corresponding to 3.224\,mm, 3.156\,mm, 2.855\,mm, and 2.802\,mm, respectively. The reconstruction of the interferometric data has been done with the Solar ALMA Pipeline (SoAP). A detailed description of the pipeline will be provided in a forthcoming publication (Szydlarski et al., in prep.; see also \cite{2020A&A...635A..71W}). One of the modes of SoAP allows to create a full-band time sequence in which all the four sub-bands are combined into one. Because of the higher signal-to-noise ratio (as a result of better sampling of Fourier space during the image reconstruction) of the full-band time series, this mode of SoAP was utilized to reconstruct the data used for the present work (instead of the four sub-bands). 

The data consists of 3 scans, 10 minutes long each, with 2~s cadence. There are gaps of approximately 140~s between the scans used for calibration measurements. The column of precipitable water vapour (PWV) in the Earth's atmosphere during the observation was 0.4\,mm, on average. The pixel size is chosen during the reconstruction process; for this observation, the sampling resolution is 0.34 arcsec. The synthetic elliptical median beam size (which defines the spatial resolution of the observations) depends on the observed frequency and, therefore, changes depending on the sub-bands or full-band used. For the full-band, the synthetic elliptical median beam size which is representative of the whole observation is 2.21 arcsec $\times$ 1.70 arcsec (1607 km $\times$ 1235 km) with an anticlockwise inclination angle of 54.06$^\circ{}$ with respect to the solar north. In addition, observations from the Solar Dynamic Observatory (SDO), consisting of photospheric magnetograms from the Helioseismic and Magnetic Imager (HMI, \cite{2012SoPh..275..207S}), and images at 1700 \AA~and 304 \AA~from the Atmospheric Imaging Assembly (AIA, \cite{2012SoPh..275...17L}) have been spatially co-aligned with the ALMA observations in order to gain context information. The co-alignment was carried out by comparing the ALMA time-averaged image for the whole observation with a time-averaged combined image of SDO/HMI, SDO/AIA1600, SDO/AIA1700 and SDO/AIA304 in the same time range. The initial point of comparison was taken from the coordinates provided by ALMA.

\begin{figure}[!t]
\centering\includegraphics[width=\textwidth]{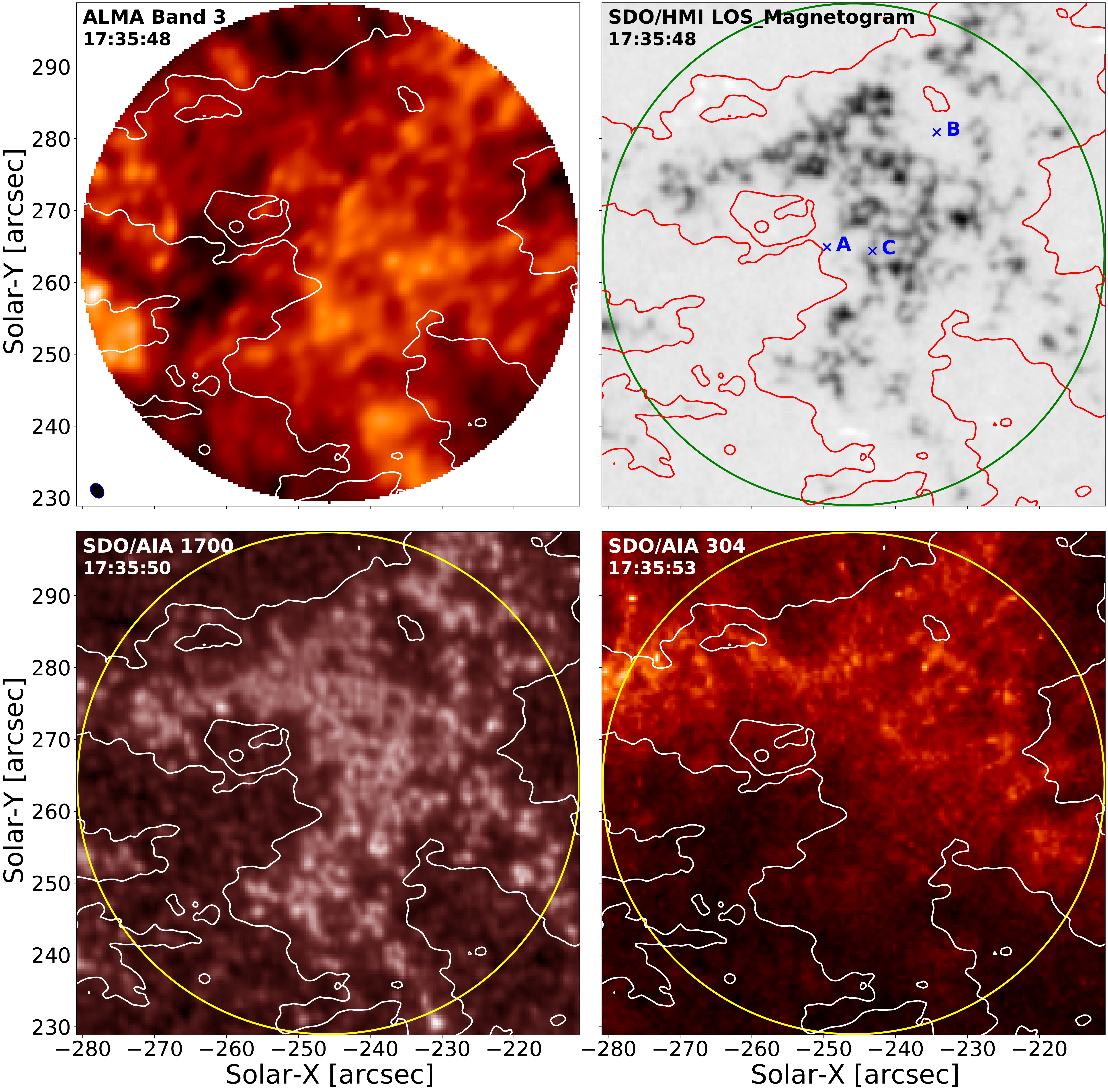}
\caption{Top left: a full-band ALMA Band-3 image at 17:35:48 UT. The small ellipse in the bottom left of the panel illustrates the synthetic elliptical median beam. Top right: line-of-sight magnetogram from SDO/HMI. Bottom left and right: SDO/AIA 1700 \AA~and SDO/AIA 304 \AA, respectively. The white/red contours delimit the network-internetwork regions and the yellow/green circles mark the ALMA field-of-view. The observing time of each image has been indicated on its top-left corner. The locations of the three small bright features of interest are labeled with A, B, and C.}
\label{fig_alma_sdo}
\end{figure}

Figure \ref{fig_alma_sdo} shows a full-band ALMA Band 3 image recorded at 17:35:48 UT (top-left), along with its co-aligned SDO/HMI line-of-sight magnetogram (top-right), SDO/AIA 1700 \AA~(bottom-left), and SDO/AIA 304 \AA~(bottom-right). In the ALMA panel, the small ellipse in the bottom-left corner illustrates the beam size. In the magnetogram panel, the blue crosses indicate the locations of three different transient brightenings whose oscillatory behaviors are studied in this work. Each of these brightenings corresponds to one single feature (labeled with A, B, and C) whose characteristics are described in Section~\ref{sec:analysis}. These ALMA observations where taken in a plage/enhanced-network region on the east side of NOAA AR12651, with strong photospheric magnetic fields, spanning a range of [-1819,961] Gauss during our observation. In all panels in Figure \ref{fig_alma_sdo}, the white/red contours depict the network-internetwork boundaries, identified from a combination of the saturated SDO/HMI magnetogram and the 1700 \AA\ image. The circles mark the FOV of ALMA on the SDO images.

\section{Method\label{sec:method}}

Characterising the structure and dynamics of magnetic features in the solar chromosphere, namely, time variations of intensity, size, and displacement, can reveal the oscillatory behaviour in this atmospheric region. Moreover, if there exists information about the magnetic-field configuration, it is also possible to inspect whether or not the oscillatory signals are magnetic in nature. In this work, three distinct bright features, one per ALMA scan (see Section~\ref{sec:obs}), are analyzed. The spatial locations of the three features, labelled  A, B, and C, are marked on the line-of-sight photospheric magnetogram in Figure~\ref{fig_alma_sdo}. We note that the magnetogram was shown for one particular frame of the observations, whereas the three features were identified at different times throughout the image sequence. However, their locations within the network region suggests their magnetic origin. A one-to-one comparison of these bright features and the magnetic elements in the HMI magnetogram is not straightforward due to the large height difference between the atmospheric layers mapped by ALMA~Band~3 and the HMI magnetogram (i.e., the high chromosphere versus the low photosphere, respectively), as well as, their relatively low spatial resolutions.

\subsection{Feature Detection\label{subsec:selection}}

Detection and tracking of small-scale bright features in the ALMA Band 3 time-series of images is not trivial, as many of the structures have dimensions comparable to the spatial resolution of the observations (or even smaller). This limits us to features larger than the beam size (i.e., larger than
$\sim 1.9$~arcsec). Several feature-tracking approaches have been developed to address similar situations. Particularly, \citet{1996JCIS..179..298C} described an algorithm to precisely identify, track, and extract quantitative information for colloidal suspensions in noisy image sequences, where large background brightenings may also exist. This is a similar challenge that we face in detecting small chromopsheric brightenings in ALMA images. The algorithm has been designed to have a high sub-pixel accuracy in locating the features. According to an error analysis by \citet{1996JCIS..179..298C}, a precision better than 0.05 pixel can be achieved. However, accounting for the rapid size and intensity variations of the small dynamic features in the solar chromosphere, we consider an (overestimated) uncertainty of about 0.5 pixel. This algorithm has been implemented in \texttt{Python} for finding blob-like features in time series of images by means of the \texttt{TrackPy} package \cite{dan_allan_2014_12255}. The algorithm has been used in various fields, including solar physics
(e.g.,\cite{2013A&A...549A.116J,2018SoPh..293..123K}). In this study, the \texttt{TrackPy} package is used to perform the tracking analysis.

By using \texttt{Trackpy}, it is possible to identify and track blob-like structures which persist in time. The main input for \texttt{Trackpy} is  an initial estimate of the size (or diameter), in pixels, of the features of interest, with which the algorithm searches for features within circles with a diameter larger than the input size. The minimum input size was chosen based on two steps: $i$) a visual inspection of the time series to identify an approximate size of transient bright features of interest, which was found to be 3~arcsec or larger; corresponding to a minimum of 8 pixels and ii) we considered the median value of the beam size (i.e., 1.93~arcsec; 6 pixels). With these in mind, the sub-pixel precision function of \texttt{TrackPy} was used to evaluate the best initial estimate size, being this 15 pixels. This is larger by a factor of $\approx2.5$ than the median beam size (i.e., the spatial resolution) and thus ensures the detection of spatially-resolved features.
The algorithm was then ran to track potential candidates in each ALMA scan, which were visually inspected in afterwards to select the three features that are analysed here. Particularly, the algorithm did not retain information on the candidate feature from frame-to-frame, meaning that an event must be present in every single frame to be considered a reliable and unique feature. Thus, the initial detection of a feature is defined as the frame in which it is first observed by the algorithm. The feature is considered extinguished when its intensity goes below the detection limit. Features that merge with other structures, or split into several features, are excluded. In addition, features with lifetimes shorter than 200~s have been excluded too. This decision was made based on a primary exploration, from which a clear manifestation of oscillatory periods on the order of 50~s or longer was noted. The features A, B, and C have lifetimes of 316~s, 238~s, and 389~s, respectively. These three selected features are being analysed in detail in the present work. An statistical and in-depth study of a larger sample is the subject of a forthcoming article.

\section{Analysis and Results\label{sec:analysis}}

The \texttt{Trackpy} procedure was applied to the identified features to extract their physical properties. Thus, for each feature at every time step, the properties, namely, size, location, and brightness temperature were determined. The size corresponds to the diameter of a circle with the same area of the feature. The location of each feature is described by its centre-of-gravity of intensity (i.e., brightness temperature). The representative brightness temperature of the feature has been chosen to be the value of the closest pixel to the centre of gravity of intensity. Also, the transverse (horizontal) velocities in solar-X and solar-Y directions (for the centre of gravity of intensity) between two consecutive frames were determined (i.e., the instantaneous horizontal velocities). All time series corresponding to these quantities were detrended (by subtracting a simple linear fit) and apodised (using a Tukey window, with a length of 0.1) prior to wavelet analysis (see Section~\ref{subsec:wavelets}).

\begin{figure}[!t]
\centering\includegraphics[width=\textwidth]{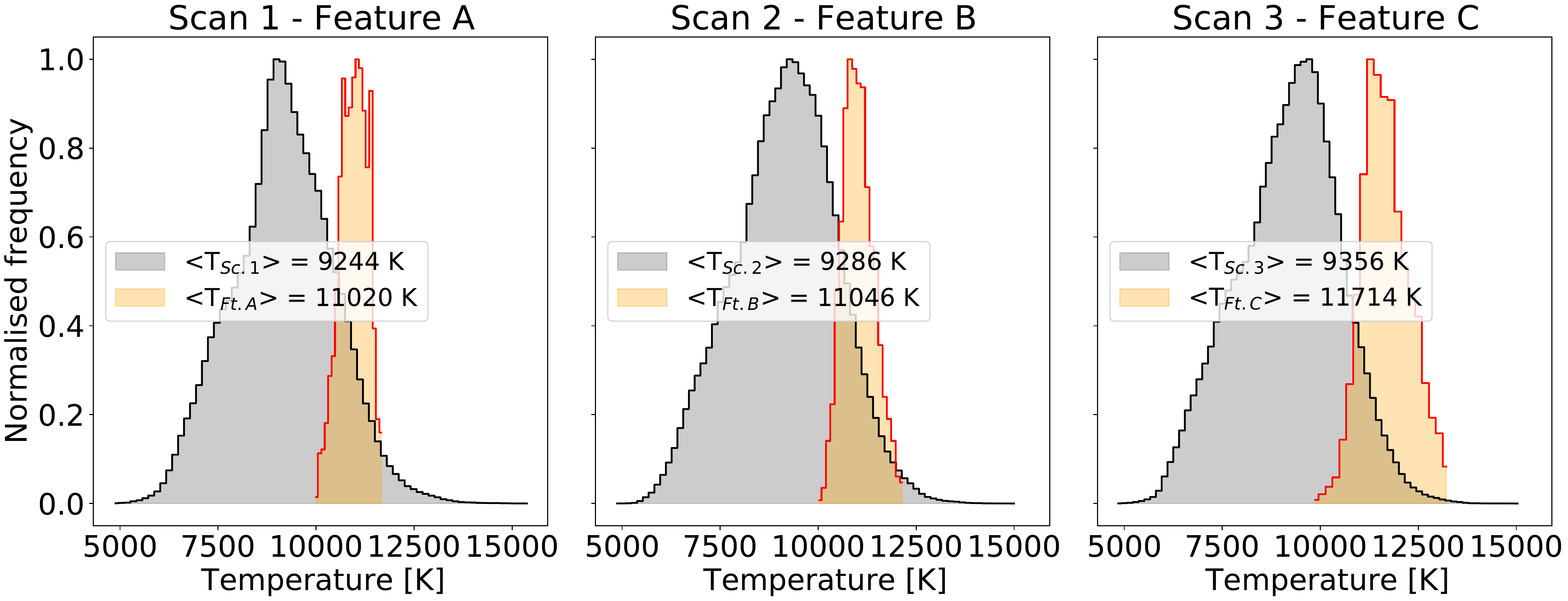}
\caption{Distributions of brightness temperature of the individual scans are shown in grey. They include the entire FOV for all time steps within the corresponding scan. The orange histogram in each panel corresponds to the temperature distribution of each identified feature during the course of its lifetime. The median brightness temperatures are given for all distributions.}
\label{fig_tempedist}
\end{figure}

\subsection{Size, location, and brightness temperature}

\begin{figure}[!t]
\centering\includegraphics[width=\textwidth]{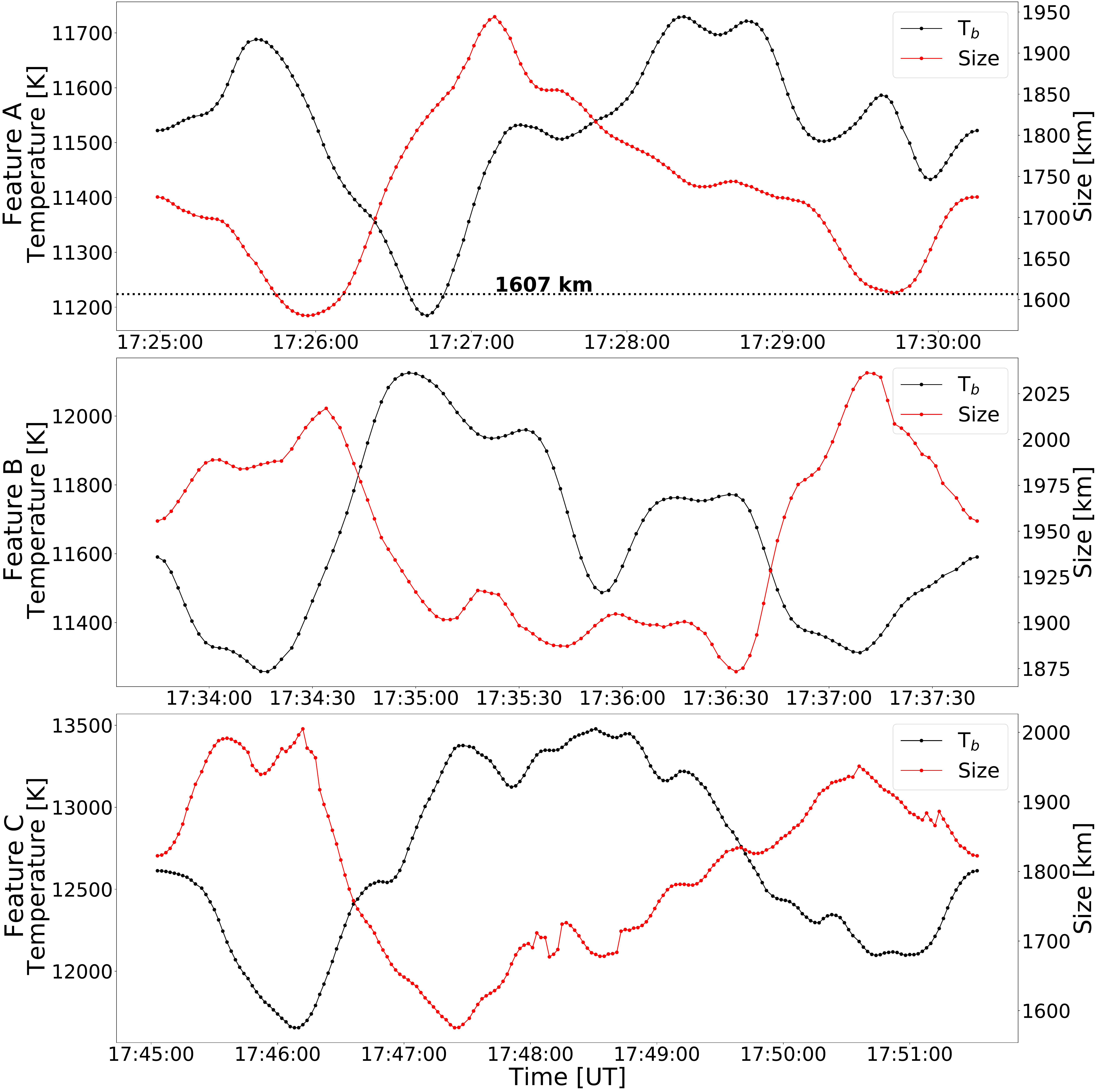}
\caption{Fluctuations in brightness temperature (black) and size (red) of the three small bright features studied in the present work. The horizontal black, dotted line in the top panel marks the size of the major axis of the synthetic elliptical median beam (i.e., the spatial-resolution element). An anti-correlation between oscillations of the two quantities is evident.}
\label{fig_tb_wd_curves}
\end{figure}
Figure \ref{fig_tempedist} shows the distributions of brightness temperature of the entire FOV during the whole observation for each scan in grey. The orange histograms correspond to the distributions of brightness temperature for the three features during their lifetimes. The median temperatures of the features A, B, and C are 1776~K, 1760~K, and 2358~K higher than the median temperatures for the entire FOV, respectively.

Figure \ref{fig_tb_wd_curves} illustrates the temporal evolution of the brightness temperature of the features in black and the temporal evolution of their sizes in red. The major axis of the synthetic elliptical median beam is plotted as a horizontal black dotted line in the first panel. It is clear that the three features under study are all spatially resolved in our observations. An anti-correlation between oscillations in the brightness temperature and size of the three features is evident. Time lags between the anti-correlations are also observed, e.g., in the second major extrema in the top panel (i.e., Feature A). Anti-phase oscillations between intensity and size of magnetic structures are suggestive signatures of fast sausage-mode waves \cite{2013A&A...555A..75M}. A quantitative inspection of the anti-phase behaviour will be described in Section~\ref{sec:analysis}\ref{subsec:wavelets}.

\subsection{Horizontal velocity\label{subsec:traject}}

\begin{figure}[!t]
\centering\includegraphics[width=\textwidth]{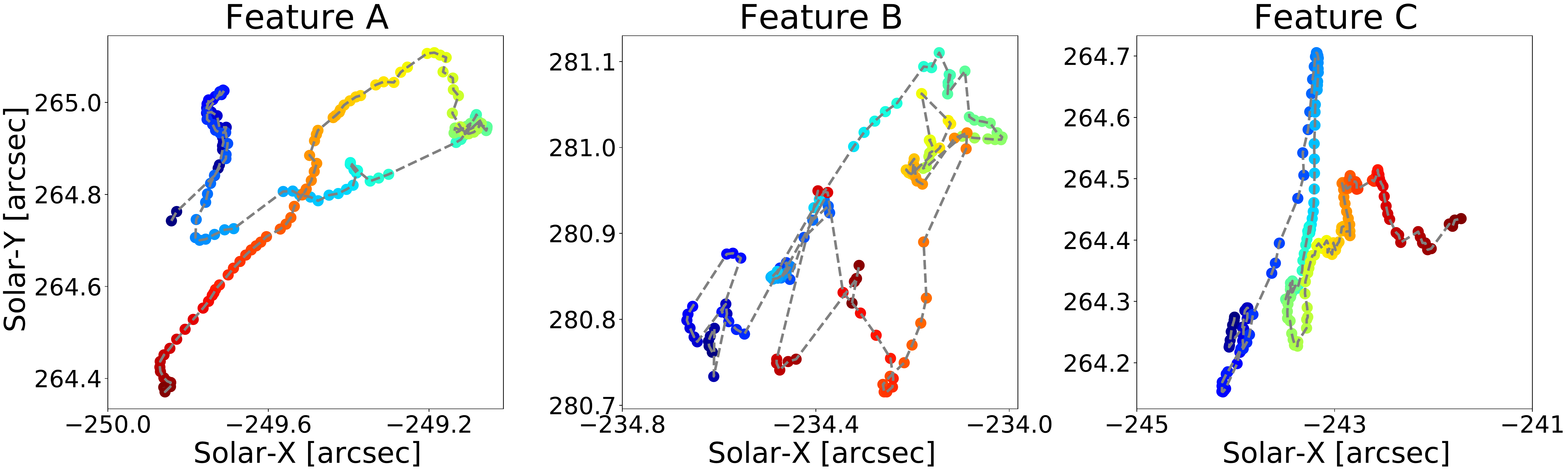}
\caption{Trajectories of the three features studied here. The colours indicate the time progression from the birth time of the features (dark blue) to when they extinguish (dark red).}
\label{fig_cm_traj}
\end{figure}

\begin{figure}[!t]
\centering\includegraphics[width=\textwidth]{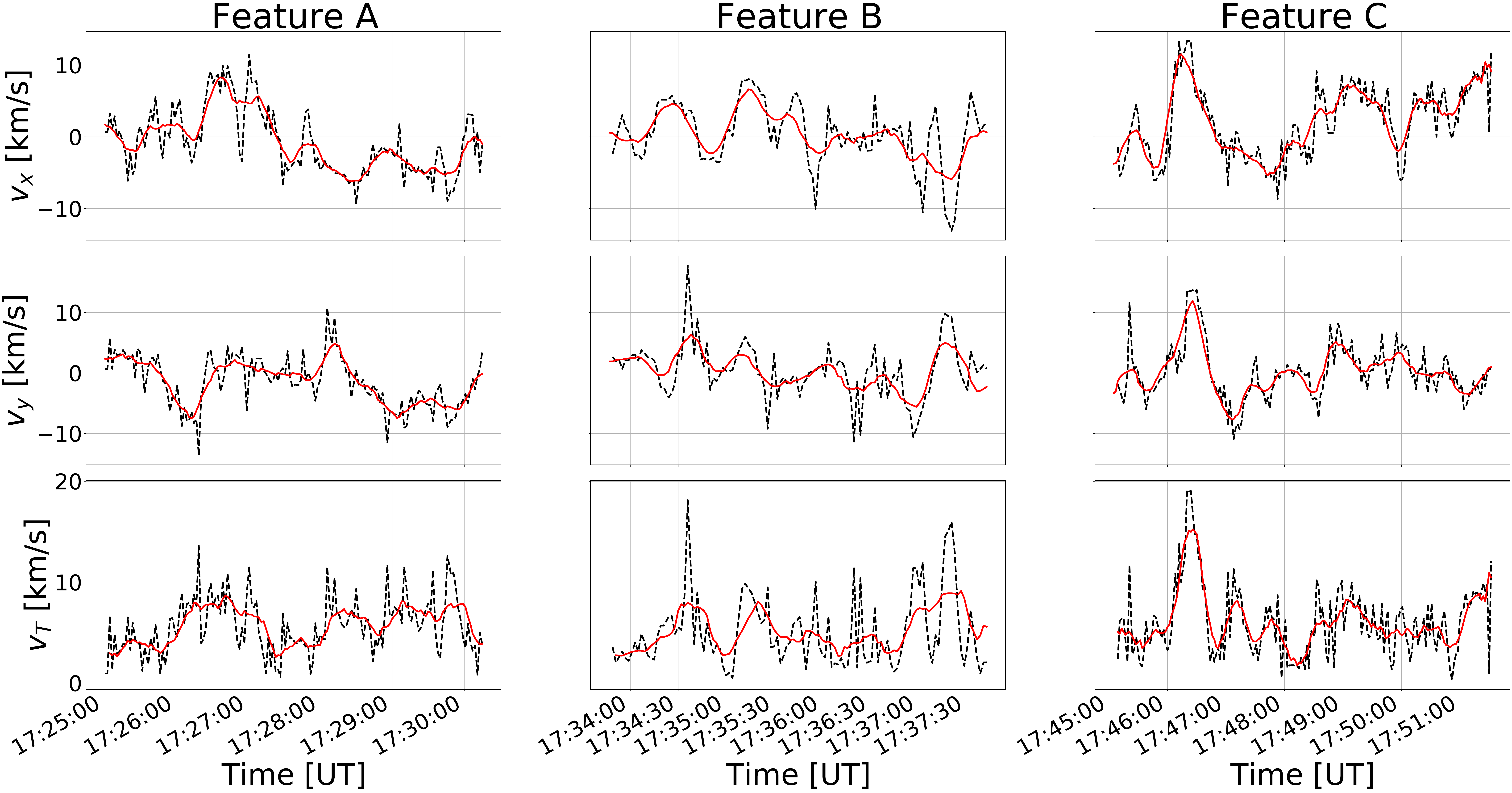}
\caption{Oscillations in horizontal velocities for the features A, B and C. In each column (associated to one feature), $v_{x}$ is shown in the top panel, $v_{y}$ in the middle panel, and the total horizontal-velocity in the bottom. The red lines are the boxcar averages of 20 seconds.}
\label{fig_cm_vels}
\end{figure}

The instantaneous horizontal velocities in each direction, i.e., $v_x$ and $v_y$, were also determined by taking the difference between the location of the centre of gravity of intensity in two consecutive frames and dividing it by 2~s, i.e., the cadence of the observations. Figure \ref{fig_cm_traj} shows trajectories of the centre of gravity of the three features under study. The colours code the time progression from dark blue to dark red. The total instantaneous horizontal velocity is then calculated as $v_T = \sqrt{v_x^2  + v_y^2}$. Figure \ref{fig_cm_vels} shows in black dashed lines the temporal evolution of the velocities $v_x$, $v_y$, and $v_T$ for the three features. The average frame to frame displacements of the gravity centres of the features are of the order of 6-9\% of the pixel size of the ALMA observations. Therefore, in order to estimate a time span for which the displacements are at least half of the pixel size, implying that the uncertainty of the measurement is reduced, the total average displacements in a mobile window of 20 seconds, i.e., adding up all the displacements frame to frame in a range of 20 seconds, have also been calculated falling in the range of 60-90\% of the pixel size. Hence, a boxcar running average of 20 seconds has been applied to the  velocities of the features and is over-plotted as red lines on all the panels in Figure~\ref{fig_cm_vels}. Moreover, in order to examine the validity of the signals (i.e., the boxcar-averaged oscillations), a statistical significance of the total velocities was performed by using the Monte Carlo (randomisation) test described by \citet{1985AJ.....90.2317L} and \citet{2001A&A...368.1095O}. In this method, the global wavelet spectrum of the boxcar averaged total velocity time-series for each feature is compared with the global wavelet spectrum obtained from a random permutation of the values in the time series. Doing this process for $N$ times, the probability $p$ with which the periodicities were produced by noise (and not a real signal) would be $M/N$, where $M$ is the number of times that the power at each frequency bin in the permuted signal is larger than the power measured from the original time series. The confidence level of the calculated periodicities from the original signal is then estimated as $1-p$. The confidence levels of the total velocity periods found with the randomisation test with $N=10000$ were $0.97$, $0.99$, and $0.99$ for features A, B, and C, respectively, meaning that they are not noise originated. Table~\ref{tab_vel} summarises the minimum, mean, median, and maximum values of the boxcar averaged total velocity amplitudes for the three features.

\begin{table}[!t]
\centering
\caption{The minimum, mean, median and maximum boxcar averaged total horizontal-velocity amplitudes for the three features under study.}
\label{tab_vel}
\begin{tabular}{lc|c|c|c}
                    & \textbf{v$_\textbf{min}$ [kms$^{-1}$] } & \textbf{v$_\textbf{mean}$ [kms$^{-1}$] } & \textbf{v$_\textbf{median}$~[kms$^{-1}$]~} & \textbf{v$_\textbf{max}$ [kms$^{-1}$] }  \\
\textbf{Feature A}  & 1.0                                     & 2.0                                      & 1.9                                        & 3.0                                      \\ 
\hline
\textbf{Feature B}  & 1.1                                     & 2.1                                      & 2.6                                        & 2.6                                      \\ 
\hline
\textbf{Feature C}  & 0.6                                     & 2.6                                      & 2.4                                        & 5.9                                      \\
\hline
\end{tabular}
\vspace*{-4pt}
\end{table}

\subsection{Wavelet analysis\label{subsec:wavelets}}

\begin{figure}[!t]
\centering\includegraphics[width=\textwidth]{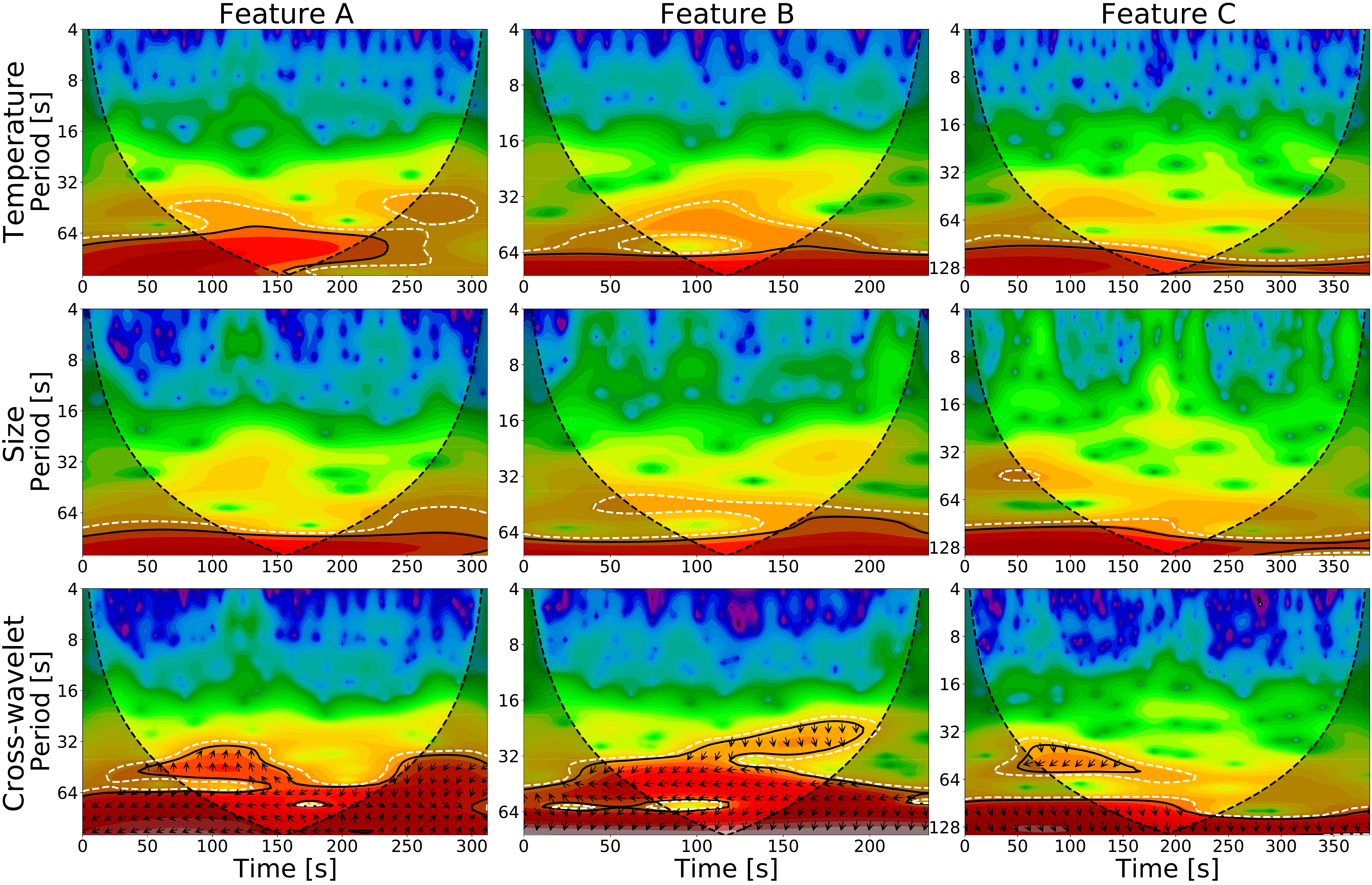}
\caption{Wavelet power spectra for oscillations in brightness temperature (top row) and size (middle row) of the three features A-C from left to right, respectively. The cross-wavelet power spectrum between the two power spectra of each feature is shown in the bottom row. In each panel, the power was plotted using a $\log_2$ scale to enhance the contrast of colours. In each panel, the $y$-axis corresponds to the oscillation period, and the $x$-axis represents the observation time of each feature in seconds. The hashed black cone region represents the cone of influence for each spectrum and the black solid contours mark the 95\% confidence levels. The white dashed contours mark the 75\% confidence level. The arrows depicted on the cross-wavelet power spectra mark phase lags between the oscillations in brightness temperature and size (arrows pointing straight down represent anti-phase relationship; pointing left indicate perturbations of brightness temperature follows that of size).}
\label{fig_Tb_Wd_wavelets}
\end{figure}

To determine the periods of the oscillation for temperature, size and velocity, as well as the phase lags between them, we perform a  wavelet analysis. \citet{1998BAMS...79...61T} presented a comprehensive description of the method which is used as the basis for this analysis. 
Accordingly, the time-varying signals are decomposed into time-frequency space components making it possible to determine the dominant oscillation modes in a power spectrum. Moreover, it is also possible to calculate the cross-wavelet spectrum between two signals which in turn gives information about their correlations, hence, the phase lags between the two signals (cf, e.g., \cite{2017ApJS..229....7G}).

\begin{table}[!h]
\centering
\caption{The range, mean and median periods of oscillation of the temperature, size and total velocity are presented for the 3 features.\label{tab_periods}}
\setlength{\tabcolsep}{.4em}
\begin{tabular}{lllllllllllll}
\multicolumn{1}{c}{} & \multicolumn{1}{c}{} & \multicolumn{3}{c}{\textbf{Temperature Periods [s]} }                             & \multicolumn{1}{c}{} & \multicolumn{3}{c}{\textbf{Size Periods [s]} }                                    & \multicolumn{1}{c}{} & \multicolumn{3}{c}{\textbf{Velocity Periods [s]} }                                 \\
\multicolumn{1}{c}{} & \multicolumn{1}{c}{} & \multicolumn{1}{c}{range} & \multicolumn{1}{c}{mean} & \multicolumn{1}{c}{median} & \multicolumn{1}{c}{} & \multicolumn{1}{c}{range} & \multicolumn{1}{c}{mean} & \multicolumn{1}{c}{median} & \multicolumn{1}{c}{} & \multicolumn{1}{c}{range} & \multicolumn{1}{c}{mean} & \multicolumn{1}{c}{median}  \\ 
\hline
\textbf{Feature A}   &                      & 60 - 114                  & 77                       & 76                         &                      & 85 - 114                  & 96                       & 96                         &                      & 40 - 108                  & 69                       & 72                          \\ 
\hline
\textbf{Feature B}   &                      & 64 - 85                   & 72                       & 72                         &                      & 64 - 85                   & 74                       & 72                         &                      & 30 - 85                   & 46                       & 43                          \\ 
\hline
\textbf{Feature C}   &                      & 102 - 136                 & 118                      & 120                        &                      & 96 - 136~                 & 115                      & 114                        &                      & 34 - 136                   & 65                       & 64                          \\
\hline
\end{tabular}
\vspace*{-4pt}
\end{table}

For each of the three features (in each column), Figure~\ref{fig_Tb_Wd_wavelets} presents the wavelet power spectra for the oscillations in brightness temperature (top row) and size (middle row) shown in Figure~\ref{fig_tb_wd_curves}, as well as the cross-wavelet power spectrum between these two quantities (bottom row). For each case, only periods which fall outside the cone of influence (the hashed areas) and the 95\% confidence-levels (being significant at 5\%; the solid contours) are considered to be representative of the oscillations. The cone of influence excludes those periods which are subject to edge effects. From the figure, a wide range of (relatively short) periods are found, of which, those within 60-136~s (for the three features) are significant at 5\%. Only these values enter in Table~\ref{tab_periods}, where the statistics of all periods extracted from the wavelet spectra for the three types of oscillations are summarised. For comparison, the white dashed contours in Figure~\ref{fig_Tb_Wd_wavelets} represent the 75\% confidence level. Periods in the ranges of 34-136~s for temperatures and 40-136~s for sizes (for features A and B) fall within this confidence level, suggesting the possible existence of shorter period oscillations in the ALMA data which will be explored in a forthcoming publication. In the present work, only period and phase-angle values within contours of 95\% confidence are analysed.
\begin{figure}[!t]
\centering\includegraphics[width=\textwidth]{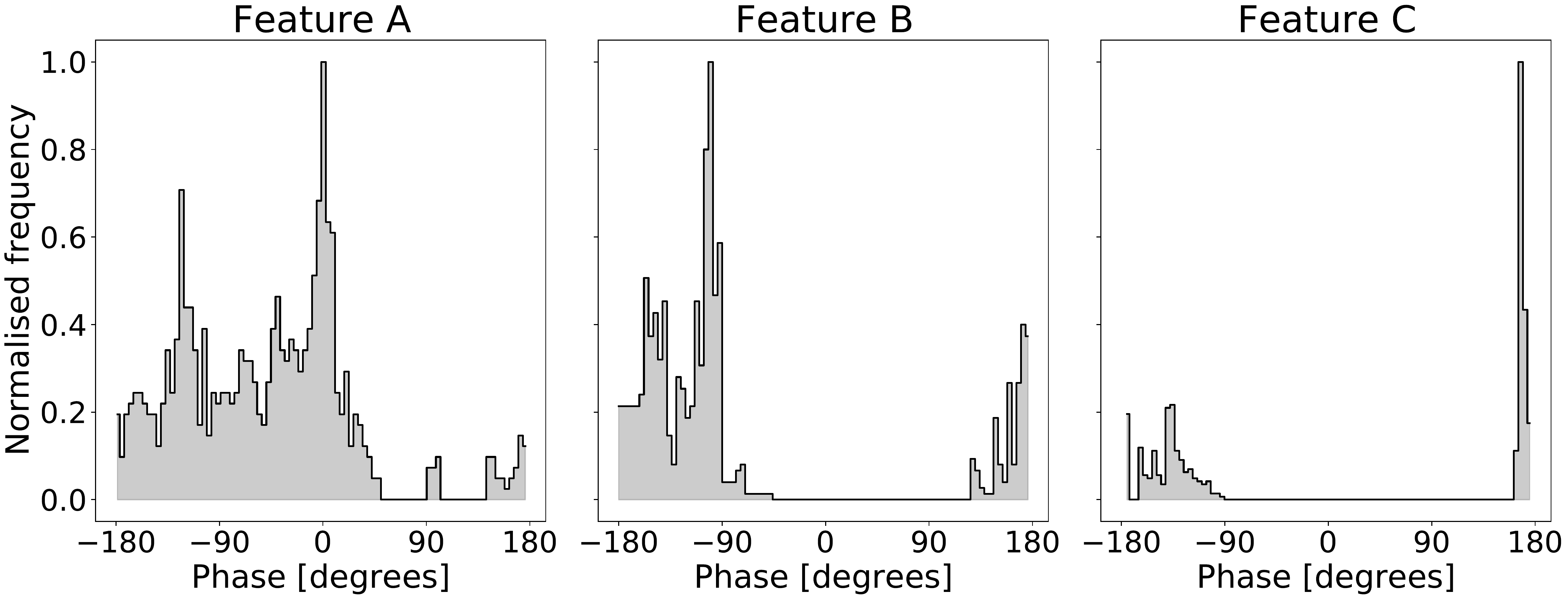}
\caption{Distribution of phase lags (in degrees) between the oscillations in brightness temperature and size for each feature.}
\label{fig_Tb_Wd_phases}
\end{figure}

In the third row of Figure~\ref{fig_Tb_Wd_wavelets}, the small black arrows indicate the phase lags between the two oscillations, being in phase when the arrows point upward and anti-phase when they point downward. If the arrows point to the right then the temperature perturbations leads that of size, and vice versa. Hence, it is possible to confirm that the features C and B show a strong anti-phase behavior for (some of) their dominant, correlated periods of oscillation while other phase differences are also evident in the case of Feature A. 
Figure~\ref{fig_Tb_Wd_phases} shows the distributions of the phase lags for each of the three features extracted from the cross-wavelet spectra (i.e., those outside the cone of influence and a 95\% confidence level). Again, features B and C present a clear anti-phase as their phase lags are mostly populated around $\pm 180^{\circ}$. For Feature A, although the phase lags are distributed over a wider range, there is also a considerable portion of the oscillations in anti-phase relationships. Other phase lags are also observed for the three features. Whether there are any differences in the nature of the oscillations observed in the three features, in particular, between the fluctuations in Feature A and those in B and C, further information about, e.g., their magnetic fields, would be required.

\begin{figure}[!tp]
\centering\includegraphics[width=\textwidth]{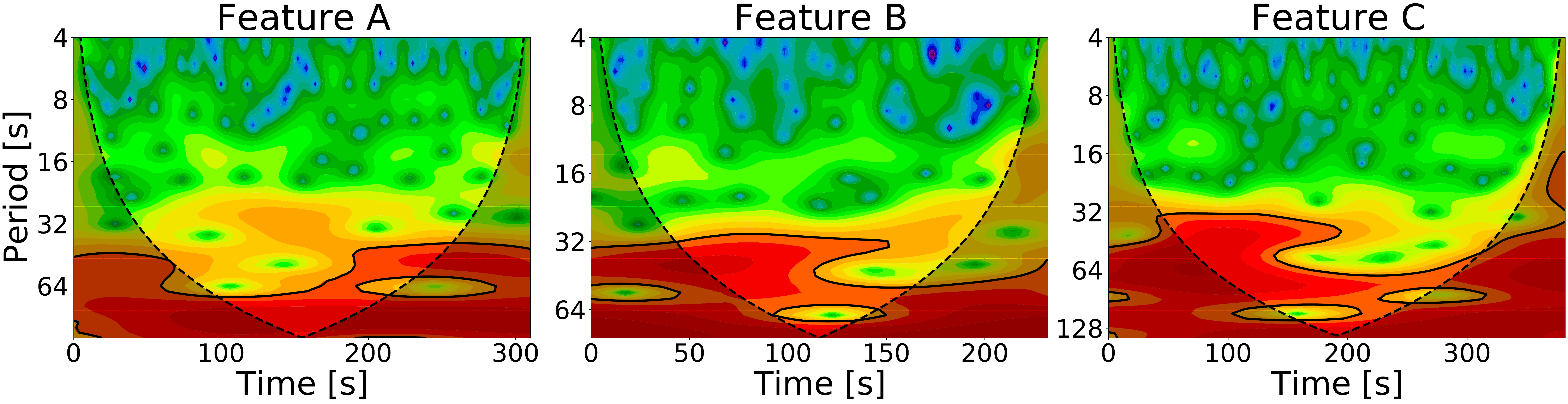}
\caption{Wavelet power spectra of the oscillations in boxcar averaged total horizontal velocities ($v_T$) for features A-C from left to right, respectively. The solid black contours mark the 95\% confidence levels and the hashed areas indicate the cone of influence (excluding regions subject to edge effect).}
\label{fig_Vt_wavelets}
\end{figure}

Figure \ref{fig_Vt_wavelets} shows the wavelet spectra of the boxcar averaged total horizontal velocity for each of the three features. The cone of influence and the 95\% confidence levels are marked with the hashed areas and solid black contours, respectively. In comparison to the brightness temperature and size, shorter periods dominate the horizontal-velocity oscillations. Nevertheless, there also exist longer periods within the significance levels, compared to those for fluctuations in the other two parameters. The statistics (providing minimum, mean, median, and maximum values of the periods of velocity oscillations) are summarised in Table~\ref{tab_periods}. The three features have an average period on the order of 66 seconds for the perturbations in horizontal velocity.

\section{Discussion and Conclusions\label{sec:dis_conc}}
\vskip6pt
We have analysed three small chromospheric bright features (transient brightenings) in ALMA Band 3 observations of a plage/enhanced-network region, revealing short-period oscillations in their brightness temperatures, sizes, and horizontal velocities. Limited by the observation's beam size (of about 1.93~arcsec), the smallest fully (spatially) resolved features under study benefited from the high quality and high temporal-resolution (i.e., 2~s cadence) of the ALMA observations. These advances resulted in reliable characterisations of the identified oscillations by means of a wavelet analysis. However, a relatively low sampling-resolution of the observations (i.e., the pixel size) resulted in some uncertainties in locating the bright features (see  Section~\ref{sec:analysis}\ref{subsec:traject}), thus, this should be taken into account when the fluctuations in the horizontal velocities are interpreted.

Clear oscillatory behaviours are observed in both brightness temperature and size of the three features under study (see Figure~\ref{fig_tb_wd_curves}). Interestingly, the fluctuations in these two parameters show an anti-phase correlation. This suggests the identification of fast sausage modes in the small chromospheric bright features. Periods of both oscillations, as well as the phase lags between them were then identified through wavelet analysis. Thus, the oscillation periods were found to be, on average, $90\pm22$~s and $110\pm12$~s for the brightness temperature and size, respectively. Similar oscillatory behaviours and periods have been reported in observations at other wavelengths in distinct chromospheric structures, such as small bright points, fibrils, spicules and pores (e.g., \cite{2011ApJ...729L..18M,2017ApJS..229....7G,2015SSRv..190..103J}). However, oscillations in small bright points were mostly observed in the low chromosphere \cite{2017ApJS..229...10J}. We expect that the features presented here (from ALMA Band 3) are formed higher in the chromosphere. It is also worth noting that the elongated structures reported by \citet{2017ApJS..229....7G} (with similar wave characteristics) have a different nature compared to the small bright features analysed here. The former represent waves in highly inclined magnetic-field lines, whereas the latter (i.e., the features studied here) are likely manifestations of cross sections of nearly vertical flux concentrations.

In addition, the horizontal displacements for all the three features are qualitatively identified in the trajectories of their centres of gravity of intensity (see Figure~\ref{fig_cm_traj}). Their oscillations in both $x$ and $y$ directions, as well as in the total horizontal velocity are evident in Figure~\ref{fig_cm_vels}. In addition, a preliminary analysis of the phase lags between the horizontal velocities and the temperature shows a predominant $\pm 90^{\circ}$ phase for Features B and C, a pattern which has been previously associated to the kink-mode nature of the transverse oscillations \cite{2013A&A...554A.115S}. However, detailed analysis of this phase relationship will be matter of future work. Furthermore, a quantitative determination of periods of the oscillations of the three features we studied revealed an average period of $66\pm23$~s for the horizontal-velocity perturbations and an average horizontal-velocity amplitude of $2.3\pm1.4$~km\,s$^{-1}$ among the three features (where the uncertainties are the standard deviations of the distributions), during the course of their lifetimes. We note that due to the large uncertainties in measuring the displacements between consecutive frames, the oscillations of the smoothed signals over 20~s have been analysed. Short periods, in the order of those found here for the transverse velocities, have been also observed in various fibrillar structures (10-500~s; \cite{2011ApJ...736L..24O,2013ApJ...764..164S,2014ApJ...784...29M,2013ApJ...768...17M,2017ApJS..229....9J}) and in bright points (43-74~s; e.g., \cite{2017ApJS..229...10J}) through the solar chromosphere, though the fibrillar structures have a wider range of periods. We should note that we cannot exclude the presence of longer periods in such small bright features presented here, hence, an statistical study of these features will be required (that is the subject of a forthcoming paper). Other values for the velocity amplitudes and the oscillation periods related to kink waves that have been reported in the literature for different observed structures in the chromosphere, span the ranges 5-29~km\,s$^{-1}$ and 37-350~s for spicules \cite{2007Sci...318.1574D,2009ApJ...705L.217H,2012ApJ...744L...5J,2012ApJ...759...18P}, 1-10~km\,s$^{-1}$ and 100-250~s \cite{2013A&A...549A.116J,2013A&A...554A.115S} for bright points, 8-11~kms$^{-1}$ and 120-180~s for mottles \cite{2012ApJ...750...51K,2013ApJ...779...82K} and for fibrils 1-7~km\,s$^{-1}$ and 94-315~s \cite{2011ApJ...739...92P,2012NatCo...3.1315M}.

We speculate that the observed anti-phase oscillations
are associated to MHD fast sausage-mode waves. While the transverse oscillations in velocity may be associated to MHD kink-mode waves, it may be also possible that the displacement are due to the bulk macroscopic motions, deprecating a noise explanation as the periodicities are significant and the randomisation test showed that the velocity oscillation cannot be reproduced randomly. The high-frequency waves identified here are of particular importance because they can penetrate into the outer atmosphere, and thus contribute to the heating budget of the upper chromosphere and corona.
Further investigations using the sub-bands capabilities of ALMA might help to determine the propagation characteristics and phase speeds of these waves, and ultimately to the energy flux carried by them into the higher layers of the solar atmosphere. In addition, a systematic analysis of a larger statistically significant sample of features is necessary in order to estimate what the typical attributes of these MHD waves are when propagating throughout the solar chromosphere. Similar studies from radiative MHD simulations will eventually help to better understand the nature of these waves. 

\vskip6pt

\enlargethispage{20pt}


\dataccess{The observational data are publicly available in the ALMA Science Archive as part of project \#2016.1.00050.S.}

\aucontribute{MS, SW, and SJ performed the data reduction and post processing. JCGG performed scientiﬁc analysis, with assistance from SJ, SW, MS, BF, and PHK. JCGG drafted the manuscript. All authors read and approved the manuscript.}

\competing{The authors declare that they have no competing interests.}

\funding{This work is supported by the SolarALMA project, which has received funding from the European Research Council (ERC) under the European Union’s Horizon 2020 research and innovation programme (grant agreement No. 682462), and by the Research Council of Norway through its Centres of Excellence scheme, project number 262622.}

\ack{This paper makes use of the following ALMA data: ADS/JAO.ALMA\#2016.1.00050.S. ALMA is a partnership of ESO (representing its member states), NSF (USA) and NINS (Japan), together with NRC(Canada), MOST and ASIAA (Taiwan), and KASI (Republic of Korea), in co-operation with the Republic of Chile. The Joint ALMA Observatory is operated by ESO, AUI/NRAO and NAOJ. We are grateful to the many colleagues who contributed to developing the solar observing modes for ALMA and for support from the ALMA Regional Centres. 
We acknowledge support from the Nordic ARC node based at the Onsala Space Observatory Swedish national infrastructure, funded through Swedish Research Council grant No 2017 – 00648, and collaboration with the Solar Simulations for the Atacama Large Millimeter Observatory Network (SSALMON, http://www.ssalmon.uio.no).
SJ, MS, BF, and PHK wish to acknowledge scientific discussions with the Waves in the Lower Solar Atmosphere (WaLSA; \href{https://www.WaLSA.team}{www.WaLSA.team}) team, which is supported by the Research Council of Norway (project no. 262622) and the Royal Society (award no. Hooke18b/SCTM).
}


\bibliographystyle{rstasj.bst}
\bibliography{references.bib}

\begin{thebibliography}{65}
\providecommand{\natexlab}[1]{#1}
\providecommand{\url}[1]{\texttt{#1}}
\expandafter\ifx\csname urlstyle\endcsname\relax
  \providecommand{\doi}[1]{doi: #1}\else
  \providecommand{\doi}{doi: \begingroup \urlstyle{rm}\Url}\fi

\bibitem[{Nakariakov} et~al.(2005){Nakariakov}, {Pascoe}, and
  {Arber}]{2005SSRv..121..115N}
{Nakariakov} VM, {Pascoe} DJ, {Arber} TD.
\newblock 2005, {Short Quasi-Periodic MHD Waves in Coronal Structures}.
\newblock \emph{\ssr}, 121\penalty0 (1-4):\penalty0 115--125.
\newblock (\doi{10.1007/s11214-006-4718-8})

\bibitem[{van Doorsselaere} et~al.(2009){van Doorsselaere}, {Verwichte}, and
  {Terradas}]{2009SSRv..149..299V}
{van Doorsselaere} T, {Verwichte} E, {Terradas} J.
\newblock 2009, {The Effect of Loop Curvature on Coronal Loop Kink
  Oscillations}.
\newblock \emph{\ssr}, 149\penalty0 (1-4):\penalty0 299--324.
\newblock (\doi{10.1007/s11214-009-9530-9})

\bibitem[{de Moortel}(2009)]{2009SSRv..149...65D}
{de Moortel} I.
\newblock 2009, {Longitudinal Waves in Coronal Loops}.
\newblock \emph{\ssr}, 149\penalty0 (1-4):\penalty0 65--81.
\newblock (\doi{10.1007/s11214-009-9526-5})

\bibitem[{Jess} and {Verth}(2016)]{2016GMS...216..449J}
{Jess} DB, {Verth} G.
\newblock 2016, {Ultra-High-Resolution Observations of MHD Waves in
  Photospheric Magnetic Structures}.
\newblock \emph{Washington DC American Geophysical Union Geophysical Monograph
  Series}, 216:\penalty0 449--465.
\newblock (\doi{10.1002/9781119055006.ch26})

\bibitem[{Jess} et~al.(2015){Jess}, {Morton}, {Verth}, {Fedun}, {Grant}, and
  {Giagkiozis}]{2015SSRv..190..103J}
{Jess} DB, {Morton} RJ, {Verth} G, {Fedun} V, {Grant} SDT, {Giagkiozis} I.
\newblock 2015, {Multiwavelength Studies of MHD Waves in the Solar
  Chromosphere. An Overview of Recent Results}.
\newblock \emph{\ssr}, 190\penalty0 (1-4):\penalty0 103--161.
\newblock (\doi{10.1007/s11214-015-0141-3})

\bibitem[{Allcock} et~al.(2019){Allcock}, {Shukhobodskaia}, {Zs{\'a}mberger},
  and {Erd{\'e}lyi}]{2019FrASS...6...48A}
{Allcock} M, {Shukhobodskaia} D, {Zs{\'a}mberger} NK, {Erd{\'e}lyi} R.
\newblock 2019, {Magnetohydrodynamic waves in multi-layered asymmetric
  waveguides: solar magneto-seismology theory and application}.
\newblock \emph{Frontiers in Astronomy and Space Sciences}, 6:\penalty0 48.
\newblock (\doi{10.3389/fspas.2019.00048})

\bibitem[{Verth} and {Jess}(2016)]{2016GMS...216..431V}
{Verth} G, {Jess} DB.
\newblock 2016, {MHD Wave Modes Resolved in Fine-Scale Chromospheric Magnetic
  Structures}.
\newblock \emph{Washington DC American Geophysical Union Geophysical Monograph
  Series}, 216:\penalty0 431--448.
\newblock (\doi{10.1002/9781119055006.ch25})

\bibitem[{De Pontieu} et~al.(2007){De Pontieu}, {McIntosh}, {Carlsson},
  {Hansteen}, {Tarbell}, {Schrijver}, {Title}, {Shine}, {Tsuneta}, {Katsukawa},
  {Ichimoto}, {Suematsu}, {Shimizu}, and {Nagata}]{2007Sci...318.1574D}
{De Pontieu} B, {McIntosh} SW, {Carlsson} M, {Hansteen} VH, {Tarbell} TD,
  {Schrijver} CJ, {Title} AM, {Shine} RA, {Tsuneta} S, {Katsukawa} Y,
  {Ichimoto} K, {Suematsu} Y, {Shimizu} T, {Nagata} S.
\newblock 2007, {Chromospheric Alfv{\'e}nic Waves Strong Enough to Power the
  Solar Wind}.
\newblock \emph{Science}, 318\penalty0 (5856):\penalty0 1574.
\newblock (\doi{10.1126/science.1151747})

\bibitem[{Lin} et~al.(2007){Lin}, {Engvold}, {Rouppe van der Voort}, and {van
  Noort}]{2007SoPh..246...65L}
{Lin} Y, {Engvold} O, {Rouppe van der Voort} LHM, {van Noort} M.
\newblock 2007, {Evidence of Traveling Waves in Filament Threads}.
\newblock \emph{\solphys}, 246\penalty0 (1):\penalty0 65--72.
\newblock (\doi{10.1007/s11207-007-0402-8})

\bibitem[{Kuridze} et~al.(2012){Kuridze}, {Morton}, {Erd{\'e}lyi}, {Dorrian},
  {Mathioudakis}, {Jess}, and {Keenan}]{2012ApJ...750...51K}
{Kuridze} D, {Morton} RJ, {Erd{\'e}lyi} R, {Dorrian} GD, {Mathioudakis} M,
  {Jess} DB, {Keenan} FP.
\newblock 2012, {Transverse Oscillations in Chromospheric Mottles}.
\newblock \emph{\apj}, 750\penalty0 (1):\penalty0 51.
\newblock (\doi{10.1088/0004-637X/750/1/51})

\bibitem[{Stangalini} et~al.(2013){Stangalini}, {Solanki}, {Cameron}, and
  {Mart{\'\i}nez Pillet}]{2013A&A...554A.115S}
{Stangalini} M, {Solanki} SK, {Cameron} R, {Mart{\'\i}nez Pillet} V.
\newblock 2013, {First evidence of interaction between longitudinal and
  transverse waves in solar magnetic elements}.
\newblock \emph{\aap}, 554:\penalty0 A115.
\newblock (\doi{10.1051/0004-6361/201220933})

\bibitem[{Stangalini} et~al.(2014){Stangalini}, {Consolini}, {Berrilli}, {De
  Michelis}, and {Tozzi}]{2014A&A...569A.102S}
{Stangalini} M, {Consolini} G, {Berrilli} F, {De Michelis} P, {Tozzi} R.
\newblock 2014, {Observational evidence for buffeting-induced kink waves in
  solar magnetic elements}.
\newblock \emph{\aap}, 569:\penalty0 A102.
\newblock (\doi{10.1051/0004-6361/201424221})

\bibitem[{Stangalini} et~al.(2015){Stangalini}, {Giannattasio}, and
  {Jafarzadeh}]{2015A&A...577A..17S}
{Stangalini} M, {Giannattasio} F, {Jafarzadeh} S.
\newblock 2015, {Non-linear propagation of kink waves to the solar
  chromosphere}.
\newblock \emph{\aap}, 577:\penalty0 A17.
\newblock (\doi{10.1051/0004-6361/201425273})

\bibitem[{Stangalini} et~al.(2017){Stangalini}, {Giannattasio}, {Erd{\'e}lyi},
  {Jafarzadeh}, {Consolini}, {Criscuoli}, {Ermolli}, {Guglielmino}, and
  {Zuccarello}]{2017ApJ...840...19S}
{Stangalini} M, {Giannattasio} F, {Erd{\'e}lyi} R, {Jafarzadeh} S, {Consolini}
  G, {Criscuoli} S, {Ermolli} I, {Guglielmino} SL, {Zuccarello} F.
\newblock 2017, {Polarized Kink Waves in Magnetic Elements: Evidence for
  Chromospheric Helical Waves}.
\newblock \emph{\apj}, 840\penalty0 (1):\penalty0 19.
\newblock (\doi{10.3847/1538-4357/aa6c5e})

\bibitem[{Jafarzadeh} et~al.(2017{\natexlab{a}}){Jafarzadeh}, {Solanki},
  {Gafeira}, {van Noort}, {Barthol}, {Blanco Rodr{\'\i}guez}, {del Toro
  Iniesta}, {Gandorfer}, {Gizon}, {Hirzberger}, {Kn{\"o}lker}, {Orozco
  Su{\'a}rez}, {Riethm{\"u}ller}, and {Schmidt}]{2017ApJS..229....9J}
{Jafarzadeh} S, {Solanki} SK, {Gafeira} R, {van Noort} M, {Barthol} P, {Blanco
  Rodr{\'\i}guez} J, {del Toro Iniesta} JC, {Gandorfer} A, {Gizon} L,
  {Hirzberger} J, {Kn{\"o}lker} M, {Orozco Su{\'a}rez} D, {Riethm{\"u}ller} TL,
  {Schmidt} W.
\newblock 2017{\natexlab{a}}, {Transverse Oscillations in Slender Ca II H
  Fibrils Observed with Sunrise/SuFI}.
\newblock \emph{\apjs}, 229\penalty0 (1):\penalty0 9.
\newblock (\doi{10.3847/1538-4365/229/1/9})

\bibitem[{Jafarzadeh} et~al.(2017{\natexlab{b}}){Jafarzadeh}, {Solanki},
  {Stangalini}, {Steiner}, {Cameron}, and {Danilovic}]{2017ApJS..229...10J}
{Jafarzadeh} S, {Solanki} SK, {Stangalini} M, {Steiner} O, {Cameron} RH,
  {Danilovic} S.
\newblock 2017{\natexlab{b}}, {High-frequency Oscillations in Small Magnetic
  Elements Observed with Sunrise/SuFI}.
\newblock \emph{\apjs}, 229\penalty0 (1):\penalty0 10.
\newblock (\doi{10.3847/1538-4365/229/1/10})

\bibitem[{Edwin} and {Roberts}(1983)]{1983SoPh...88..179E}
{Edwin} PM, {Roberts} B.
\newblock 1983, {Wave Propagation in a Magnetic Cylinder}.
\newblock \emph{\solphys}, 88\penalty0 (1-2):\penalty0 179--191.
\newblock (\doi{10.1007/BF00196186})

\bibitem[{Erd{\'e}lyi} and {Morton}(2009)]{2009A&A...494..295E}
{Erd{\'e}lyi} R, {Morton} RJ.
\newblock 2009, {Magnetohydrodynamic waves in a compressible magnetic flux tube
  with elliptical cross-section}.
\newblock \emph{\aap}, 494\penalty0 (1):\penalty0 295--309.
\newblock (\doi{10.1051/0004-6361:200810318})

\bibitem[{Morton} et~al.(2011){Morton}, {Erd{\'e}lyi}, {Jess}, and
  {Mathioudakis}]{2011ApJ...729L..18M}
{Morton} RJ, {Erd{\'e}lyi} R, {Jess} DB, {Mathioudakis} M.
\newblock 2011, {Observations of Sausage Modes in Magnetic Pores}.
\newblock \emph{\apjl}, 729\penalty0 (2):\penalty0 L18.
\newblock (\doi{10.1088/2041-8205/729/2/L18})

\bibitem[{Jess} et~al.(2010){Jess}, {Mathioudakis}, {Christian}, {Keenan},
  {Ryans}, and {Crockett}]{2010SoPh..261..363J}
{Jess} DB, {Mathioudakis} M, {Christian} DJ, {Keenan} FP, {Ryans} RSI,
  {Crockett} PJ.
\newblock 2010, {ROSA: A High-cadence, Synchronized Multi-camera Solar Imaging
  System}.
\newblock \emph{\solphys}, 261\penalty0 (2):\penalty0 363--373.
\newblock (\doi{10.1007/s11207-009-9500-0})

\bibitem[{Gafeira} et~al.(2017){Gafeira}, {Jafarzadeh}, {Solanki}, {Lagg}, {van
  Noort}, {Barthol}, {Blanco Rodr{\'\i}guez}, {del Toro Iniesta}, {Gandorfer},
  {Gizon}, {Hirzberger}, {Kn{\"o}lker}, {Orozco Su{\'a}rez}, {Riethm{\"u}ller},
  and {Schmidt}]{2017ApJS..229....7G}
{Gafeira} R, {Jafarzadeh} S, {Solanki} SK, {Lagg} A, {van Noort} M, {Barthol}
  P, {Blanco Rodr{\'\i}guez} J, {del Toro Iniesta} JC, {Gandorfer} A, {Gizon}
  L, {Hirzberger} J, {Kn{\"o}lker} M, {Orozco Su{\'a}rez} D, {Riethm{\"u}ller}
  TL, {Schmidt} W.
\newblock 2017, {Oscillations on Width and Intensity of Slender Ca II H Fibrils
  from Sunrise/SuFI}.
\newblock \emph{\apjs}, 229\penalty0 (1):\penalty0 7.
\newblock (\doi{10.3847/1538-4365/229/1/7})

\bibitem[{Solanki} et~al.(2010){Solanki}, {Barthol}, {Danilovic}, {Feller},
  {Gandorfer}, {Hirzberger}, {Riethm{\"u}ller}, {Sch{\"u}ssler}, {Bonet},
  {Mart{\'\i}nez Pillet}, {del Toro Iniesta}, {Domingo}, {Palacios},
  {Kn{\"o}lker}, {Bello Gonz{\'a}lez}, {Berkefeld}, {Franz}, {Schmidt}, and
  {Title}]{2010ApJ...723L.127S}
{Solanki} SK, {Barthol} P, {Danilovic} S, {Feller} A, {Gandorfer} A,
  {Hirzberger} J, {Riethm{\"u}ller} TL, {Sch{\"u}ssler} M, {Bonet} JA,
  {Mart{\'\i}nez Pillet} V, {del Toro Iniesta} JC, {Domingo} V, {Palacios} J,
  {Kn{\"o}lker} M, {Bello Gonz{\'a}lez} N, {Berkefeld} T, {Franz} M, {Schmidt}
  W, {Title} AM.
\newblock 2010, {SUNRISE: Instrument, Mission, Data, and First Results}.
\newblock \emph{\apjl}, 723\penalty0 (2):\penalty0 L127--L133.
\newblock (\doi{10.1088/2041-8205/723/2/L127})

\bibitem[{Solanki} et~al.(2017){Solanki}, {Riethm{\"u}ller}, {Barthol},
  {Danilovic}, {Deutsch}, {Doerr}, {Feller}, {Gandorfer}, {Germerott}, {Gizon},
  {Grauf}, {Heerlein}, {Hirzberger}, {Kolleck}, {Lagg}, {Meller}, {Tomasch},
  {van Noort}, {Blanco Rodr{\'\i}guez}, {Gasent Blesa}, {Balaguer Jim{\'e}nez},
  {Del Toro Iniesta}, {L{\'o}pez Jim{\'e}nez}, {Orozco Suarez}, {Berkefeld},
  {Halbgewachs}, {Schmidt}, {{\'A}lvarez-Herrero}, {Sabau-Graziati}, {P{\'e}rez
  Grand e}, {Mart{\'\i}nez Pillet}, {Card}, {Centeno}, {Kn{\"o}lker}, and
  {Lecinski}]{2017ApJS..229....2S}
{Solanki} SK, {Riethm{\"u}ller} TL, {Barthol} P, {Danilovic} S, {Deutsch} W,
  {Doerr} HP, {Feller} A, {Gandorfer} A, {Germerott} D, {Gizon} L, {Grauf} B,
  {Heerlein} K, {Hirzberger} J, {Kolleck} M, {Lagg} A, {Meller} R, {Tomasch} G,
  {van Noort} M, {Blanco Rodr{\'\i}guez} J, {Gasent Blesa} JL, {Balaguer
  Jim{\'e}nez} M, {Del Toro Iniesta} JC, {L{\'o}pez Jim{\'e}nez} AC, {Orozco
  Suarez} D, {Berkefeld} T, {Halbgewachs} C, {Schmidt} W, {{\'A}lvarez-Herrero}
  A, {Sabau-Graziati} L, {P{\'e}rez Grand e} I, {Mart{\'\i}nez Pillet} V,
  {Card} G, {Centeno} R, {Kn{\"o}lker} M, {Lecinski} A.
\newblock 2017, {The Second Flight of the Sunrise Balloon-borne Solar
  Observatory: Overview of Instrument Updates, the Flight, the Data, and First
  Results}.
\newblock \emph{\apjs}, 229\penalty0 (1):\penalty0 2.
\newblock (\doi{10.3847/1538-4365/229/1/2})

\bibitem[{Wootten} and {Thompson}(2009)]{2009IEEEP..97.1463W}
{Wootten} A, {Thompson} AR.
\newblock 2009, {The Atacama Large Millimeter/Submillimeter Array}.
\newblock \emph{IEEE Proceedings}, 97\penalty0 (8):\penalty0 1463--1471.
\newblock (\doi{10.1109/JPROC.2009.2020572})

\bibitem[{Wedemeyer-B{\"o}hm} et~al.(2007){Wedemeyer-B{\"o}hm}, {Steiner},
  {Bruls}, and {Rammacher}]{2007ASPC..368...93W}
{Wedemeyer-B{\"o}hm} S, {Steiner} O, {Bruls} J, {Rammacher} W.
\newblock 2007,
\newblock , \emph{The Physics of Chromospheric Plasmas}, 93volume 368 of
  \emph{Astronomical Society of the Pacific Conference Series}pages

\bibitem[{Bastian}(2002)]{2002AN....323..271B}
{Bastian} TS.
\newblock 2002, {ALMA and the Sun}.
\newblock \emph{Astronomische Nachrichten}, 323:\penalty0 271--276.
\newblock (\doi{10.1002/1521-3994(200208)323:3/4<271::AID-ASNA271>3.0.CO;2-1})

\bibitem[{Wedemeyer}(2016)]{2016Msngr.163...15W}
{Wedemeyer} S.
\newblock 2016, {New Eyes on the Sun {\textemdash} Solar Science with ALMA}.
\newblock \emph{The Messenger}, 163:\penalty0 15--20

\bibitem[{Wedemeyer} et~al.(2016){Wedemeyer}, {Bastian}, {Braj{\v{s}}a},
  {Hudson}, {Fleishman}, {Loukitcheva}, {Fleck}, {Kontar}, {De Pontieu},
  {Yagoubov}, {Tiwari}, {Soler}, {Black}, {Antolin}, {Scullion}, {Gun{\'a}r},
  {Labrosse}, {Ludwig}, {Benz}, {White}, {Hauschildt}, {Doyle}, {Nakariakov},
  {Ayres}, {Heinzel}, {Karlicky}, {Van Doorsselaere}, {Gary}, {Alissandrakis},
  {Nindos}, {Solanki}, {Rouppe van der Voort}, {Shimojo}, {Kato},
  {Zaqarashvili}, {Perez}, {Selhorst}, and {Barta}]{2016SSRv..200....1W}
{Wedemeyer} S, {Bastian} T, {Braj{\v{s}}a} R, {Hudson} H, {Fleishman} G,
  {Loukitcheva} M, {Fleck} B, {Kontar} EP, {De Pontieu} B, {Yagoubov} P,
  {Tiwari} SK, {Soler} R, {Black} JH, {Antolin} P, {Scullion} E, {Gun{\'a}r} S,
  {Labrosse} N, {Ludwig} HG, {Benz} AO, {White} SM, {Hauschildt} P, {Doyle} JG,
  {Nakariakov} VM, {Ayres} T, {Heinzel} P, {Karlicky} M, {Van Doorsselaere} T,
  {Gary} D, {Alissandrakis} CE, {Nindos} A, {Solanki} SK, {Rouppe van der
  Voort} L, {Shimojo} M, {Kato} Y, {Zaqarashvili} T, {Perez} E, {Selhorst} CL,
  {Barta} M.
\newblock 2016, {Solar Science with the Atacama Large Millimeter/Submillimeter
  Array{\textemdash}A New View of Our Sun}.
\newblock \emph{\ssr}, 200\penalty0 (1-4):\penalty0 1--73.
\newblock (\doi{10.1007/s11214-015-0229-9})

\bibitem[{Loukitcheva} et~al.(2017){Loukitcheva}, {White}, {Solanki},
  {Fleishman}, and {Carlsson}]{2017A&A...601A..43L}
{Loukitcheva} M, {White} SM, {Solanki} SK, {Fleishman} GD, {Carlsson} M.
\newblock 2017, {Millimeter radiation from a 3D model of the solar atmosphere.
  II. Chromospheric magnetic field}.
\newblock \emph{\aap}, 601:\penalty0 A43.
\newblock (\doi{10.1051/0004-6361/201629099})

\bibitem[{Shimojo} et~al.(2017{\natexlab{a}}){Shimojo}, {Bastian}, {Hales},
  {White}, {Iwai}, {Hills}, {Hirota}, {Phillips}, {Sawada}, {Yagoubov},
  {Siringo}, {Asayama}, {Sugimoto}, {Braj{\v{s}}a}, {Skoki{\'c}}, {B{\'a}rta},
  {Kim}, {de Gregorio-Monsalvo}, {Corder}, {Hudson}, {Wedemeyer}, {Gary}, {De
  Pontieu}, {Loukitcheva}, {Fleishman}, {Chen}, {Kobelski}, and
  {Yan}]{2017SoPh..292...87S}
{Shimojo} M, {Bastian} TS, {Hales} AS, {White} SM, {Iwai} K, {Hills} RE,
  {Hirota} A, {Phillips} NM, {Sawada} T, {Yagoubov} P, {Siringo} G, {Asayama}
  S, {Sugimoto} M, {Braj{\v{s}}a} R, {Skoki{\'c}} I, {B{\'a}rta} M, {Kim} S,
  {de Gregorio-Monsalvo} I, {Corder} SA, {Hudson} HS, {Wedemeyer} S, {Gary} DE,
  {De Pontieu} B, {Loukitcheva} M, {Fleishman} GD, {Chen} B, {Kobelski} A,
  {Yan} Y.
\newblock 2017{\natexlab{a}}, {Observing the Sun with the Atacama Large
  Millimeter/submillimeter Array (ALMA): High-Resolution Interferometric
  Imaging}.
\newblock \emph{\solphys}, 292\penalty0 (7):\penalty0 87.
\newblock (\doi{10.1007/s11207-017-1095-2})

\bibitem[{Shimojo} et~al.(2017{\natexlab{b}}){Shimojo}, {Hudson}, {White},
  {Bastian}, and {Iwai}]{2017ApJ...841L...5S}
{Shimojo} M, {Hudson} HS, {White} SM, {Bastian} TS, {Iwai} K.
\newblock 2017{\natexlab{b}}, {The First ALMA Observation of a Solar Plasmoid
  Ejection from an X-Ray Bright Point}.
\newblock \emph{\apjl}, 841\penalty0 (1):\penalty0 L5.
\newblock (\doi{10.3847/2041-8213/aa70e3})

\bibitem[{Yokoyama} et~al.(2018){Yokoyama}, {Shimojo}, {Okamoto}, and
  {Iijima}]{2018ApJ...863...96Y}
{Yokoyama} T, {Shimojo} M, {Okamoto} TJ, {Iijima} H.
\newblock 2018, {ALMA Observations of the Solar Chromosphere on the Polar
  Limb}.
\newblock \emph{\apj}, 863\penalty0 (1):\penalty0 96.
\newblock (\doi{10.3847/1538-4357/aad27e})

\bibitem[{Nindos} et~al.(2018){Nindos}, {Alissandrakis}, {Bastian},
  {Patsourakos}, {De Pontieu}, {Warren}, {Ayres}, {Hudson}, {Shimizu}, {Vial},
  {Wedemeyer}, and {Yurchyshyn}]{2018A&A...619L...6N}
{Nindos} A, {Alissandrakis} CE, {Bastian} TS, {Patsourakos} S, {De Pontieu} B,
  {Warren} H, {Ayres} T, {Hudson} HS, {Shimizu} T, {Vial} JC, {Wedemeyer} S,
  {Yurchyshyn} V.
\newblock 2018, {First high-resolution look at the quiet Sun with ALMA at 3mm}.
\newblock \emph{\aap}, 619:\penalty0 L6.
\newblock (\doi{10.1051/0004-6361/201834113})

\bibitem[{Jafarzadeh} et~al.(2019){Jafarzadeh}, {Wedemeyer}, {Szydlarski}, {De
  Pontieu}, {Rezaei}, and {Carlsson}]{2019A&A...622A.150J}
{Jafarzadeh} S, {Wedemeyer} S, {Szydlarski} M, {De Pontieu} B, {Rezaei} R,
  {Carlsson} M.
\newblock 2019, {The solar chromosphere at millimetre and ultraviolet
  wavelengths. I. Radiation temperatures and a detailed comparison}.
\newblock \emph{\aap}, 622:\penalty0 A150.
\newblock (\doi{10.1051/0004-6361/201834205})

\bibitem[{Loukitcheva} et~al.(2019){Loukitcheva}, {White}, and
  {Solanki}]{2019ApJ...877L..26L}
{Loukitcheva} MA, {White} SM, {Solanki} SK.
\newblock 2019, {ALMA Detection of Dark Chromospheric Holes in the Quiet Sun}.
\newblock \emph{\apjl}, 877\penalty0 (2):\penalty0 L26.
\newblock (\doi{10.3847/2041-8213/ab2191})

\bibitem[{Patsourakos} et~al.(2020){Patsourakos}, {Alissandrakis}, {Nindos},
  and {Bastian}]{2020A&A...634A..86P}
{Patsourakos} S, {Alissandrakis} CE, {Nindos} A, {Bastian} TS.
\newblock 2020, {Observations of solar chromospheric oscillations at 3 mm with
  ALMA}.
\newblock \emph{\aap}, 634:\penalty0 A86.
\newblock (\doi{10.1051/0004-6361/201936618})

\bibitem[{da Silva Santos} et~al.(2020){da Silva Santos}, {de la Cruz
  Rodr{\'\i}guez}, {Leenaarts}, {Chintzoglou}, {De Pontieu}, {Wedemeyer}, and
  {Szydlarski}]{2020A&A...634A..56D}
{da Silva Santos} JM, {de la Cruz Rodr{\'\i}guez} J, {Leenaarts} J,
  {Chintzoglou} G, {De Pontieu} B, {Wedemeyer} S, {Szydlarski} M.
\newblock 2020, {The multi-thermal chromosphere. Inversions of ALMA and IRIS
  data}.
\newblock \emph{\aap}, 634:\penalty0 A56.
\newblock (\doi{10.1051/0004-6361/201937117})

\bibitem[{Shimojo} et~al.(2020){Shimojo}, {Kawate}, {Okamoto}, {Yokoyama},
  {Narukage}, {Sakao}, {Iwai}, {Fleishman}, and {Shibata}]{2020ApJ...888L..28S}
{Shimojo} M, {Kawate} T, {Okamoto} TJ, {Yokoyama} T, {Narukage} N, {Sakao} T,
  {Iwai} K, {Fleishman} GD, {Shibata} K.
\newblock 2020, {Estimating the Temperature and Density of a Spicule from 100
  GHz Data Obtained with ALMA}.
\newblock \emph{\apjl}, 888\penalty0 (2):\penalty0 L28.
\newblock (\doi{10.3847/2041-8213/ab62a5})

\bibitem[{Wedemeyer} et~al.(2020){Wedemeyer}, {Szydlarski}, {Jafarzadeh},
  {Eklund}, {Guevara Gomez}, {Bastian}, {Fleck}, {de la Cruz Rodriguez},
  {Rodger}, and {Carlsson}]{2020A&A...635A..71W}
{Wedemeyer} S, {Szydlarski} M, {Jafarzadeh} S, {Eklund} H, {Guevara Gomez} JC,
  {Bastian} T, {Fleck} B, {de la Cruz Rodriguez} J, {Rodger} A, {Carlsson} M.
\newblock 2020, {The Sun at millimeter wavelengths. I. Introduction to ALMA
  Band 3 observations}.
\newblock \emph{\aap}, 635:\penalty0 A71.
\newblock (\doi{10.1051/0004-6361/201937122})

\bibitem[{Nindos} et~al.(2020){Nindos}, {Alissandrakis}, {Patsourakos}, and
  {Bastian}]{2020A&A...638A..62N}
{Nindos} A, {Alissandrakis} CE, {Patsourakos} S, {Bastian} TS.
\newblock 2020, {Transient brightenings in the quiet Sun detected by ALMA at 3
  mm}.
\newblock \emph{\aap}, 638:\penalty0 A62.
\newblock (\doi{10.1051/0004-6361/202037810})

\bibitem[{Eklund} et~al.(2020{\natexlab{a}}){Eklund}, {Wedemeyer},
  {Szydlarski}, {Jafarzadeh}, and {Guevara Gómez}]{Eklund2020}
{Eklund} H, {Wedemeyer} S, {Szydlarski} M, {Jafarzadeh} S, {Guevara Gómez} JC.
\newblock 2020{\natexlab{a}}, {The Sun at millimeter wavelengths II.
  Small-scale dynamic events in ALMA Band 3}.
\newblock \emph{\aap}, in preparation

\bibitem[{Eklund} et~al.(2020{\natexlab{b}}){Eklund}, {Wedemeyer}, {Snow},
  {Jess}, {Jafarzadeh}, {Grant}, {Carlsson}, and {Szydlarski}]{Eklund2020RS}
{Eklund} H, {Wedemeyer} S, {Snow} B, {Jess} DB, {Jafarzadeh} S, {Grant} SDT,
  {Carlsson} M, {Szydlarski} M.
\newblock 2020{\natexlab{b}}, {Characterisation of shock wave signatures at
  millimetre wavelengths from Bifrost simulations}.
\newblock \emph{\ptrsa}, \penalty0 (current issue)

\bibitem[{Chintzoglou} et~al.(2020){Chintzoglou}, {De Pontieu},
  {Mart{\'\i}nez-Sykora}, {Hansteen}, {de la Cruz Rodr{\'\i}guez},
  {Szydlarski}, {Jafarzadeh}, {Wedemeyer}, {Bastian}, and {Sa{\'\i}nz
  Dalda}]{2020arXiv200512717C}
{Chintzoglou} G, {De Pontieu} B, {Mart{\'\i}nez-Sykora} J, {Hansteen} V, {de la
  Cruz Rodr{\'\i}guez} J, {Szydlarski} M, {Jafarzadeh} S, {Wedemeyer} S,
  {Bastian} TS, {Sa{\'\i}nz Dalda} A.
\newblock 2020, {IRIS and ALMA Observations Uncovering a Type-II Spicule and
  the Dynamic Nature of a Chromospheric Plage Region}.
\newblock \emph{arXiv e-prints}, art. arXiv:2005.12717

\bibitem[{Jafarzadeh} et~al.(2020){Jafarzadeh}, {Wedemeyer}, {Fleck},
  {Stangalini}, , {Jess}, {Morton}, {Szydlarski}, {Henriques}, {Zhu},
  {Wiegelmann}, {Guevara Gómez}, {Grant}, {Chen}, {Reardon}, and
  {White}]{Jafarzadeh2020}
{Jafarzadeh} S, {Wedemeyer} S, {Fleck} B, {Stangalini} M, , {Jess} DB, {Morton}
  RJ, {Szydlarski} M, {Henriques} VMJ, {Zhu} X, {Wiegelmann} T, {Guevara
  Gómez} JC, {Grant} SDG, {Chen} B, {Reardon} K, {White} SM.
\newblock 2020, {An overall view of temperature oscillations in the solar
  chromosphere with ALMA}.
\newblock \emph{\ptrsa}, \penalty0 (current issue)

\bibitem[{Narang} et~al.(2020){Narang}, {Chandrashekhar}, {Jafarzadeh},
  {Wedemeyer}, {Fleck}, and {Szydlarski}]{Narang2020}
{Narang} N, {Chandrashekhar} K, {Jafarzadeh} S, {Wedemeyer} S, {Fleck} B,
  {Szydlarski} M.
\newblock 2020, {Power distribution of oscillations in a plage region observed
  with ALMA, IRIS, and SDO}.
\newblock \emph{\ptrsa}, \penalty0 (current issue)

\bibitem[{Scherrer} et~al.(2012){Scherrer}, {Schou}, {Bush}, {Kosovichev},
  {Bogart}, {Hoeksema}, {Liu}, {Duvall}, {Zhao}, {Title}, {Schrijver},
  {Tarbell}, and {Tomczyk}]{2012SoPh..275..207S}
{Scherrer} PH, {Schou} J, {Bush} RI, {Kosovichev} AG, {Bogart} RS, {Hoeksema}
  JT, {Liu} Y, {Duvall} TL, {Zhao} J, {Title} AM, {Schrijver} CJ, {Tarbell} TD,
  {Tomczyk} S.
\newblock 2012, {The Helioseismic and Magnetic Imager (HMI) Investigation for
  the Solar Dynamics Observatory (SDO)}.
\newblock \emph{\solphys}, 275\penalty0 (1-2):\penalty0 207--227.
\newblock (\doi{10.1007/s11207-011-9834-2})

\bibitem[{Lemen} et~al.(2012){Lemen}, {Title}, {Akin}, {Boerner}, {Chou},
  {Drake}, {Duncan}, {Edwards}, {Friedlaender}, {Heyman}, {Hurlburt}, {Katz},
  {Kushner}, {Levay}, {Lindgren}, {Mathur}, {McFeaters}, {Mitchell}, {Rehse},
  {Schrijver}, {Springer}, {Stern}, {Tarbell}, {Wuelser}, {Wolfson}, {Yanari},
  {Bookbinder}, {Cheimets}, {Caldwell}, {Deluca}, {Gates}, {Golub}, {Park},
  {Podgorski}, {Bush}, {Scherrer}, {Gummin}, {Smith}, {Auker}, {Jerram},
  {Pool}, {Soufli}, {Windt}, {Beardsley}, {Clapp}, {Lang}, and
  {Waltham}]{2012SoPh..275...17L}
{Lemen} JR, {Title} AM, {Akin} DJ, {Boerner} PF, {Chou} C, {Drake} JF, {Duncan}
  DW, {Edwards} CG, {Friedlaender} FM, {Heyman} GF, {Hurlburt} NE, {Katz} NL,
  {Kushner} GD, {Levay} M, {Lindgren} RW, {Mathur} DP, {McFeaters} EL,
  {Mitchell} S, {Rehse} RA, {Schrijver} CJ, {Springer} LA, {Stern} RA,
  {Tarbell} TD, {Wuelser} JP, {Wolfson} CJ, {Yanari} C, {Bookbinder} JA,
  {Cheimets} PN, {Caldwell} D, {Deluca} EE, {Gates} R, {Golub} L, {Park} S,
  {Podgorski} WA, {Bush} RI, {Scherrer} PH, {Gummin} MA, {Smith} P, {Auker} G,
  {Jerram} P, {Pool} P, {Soufli} R, {Windt} DL, {Beardsley} S, {Clapp} M,
  {Lang} J, {Waltham} N.
\newblock 2012, {The Atmospheric Imaging Assembly (AIA) on the Solar Dynamics
  Observatory (SDO)}.
\newblock \emph{\solphys}, 275\penalty0 (1-2):\penalty0 17--40.
\newblock (\doi{10.1007/s11207-011-9776-8})

\bibitem[{Crocker} and {Grier}(1996)]{1996JCIS..179..298C}
{Crocker} JC, {Grier} DG.
\newblock 1996, {Methods of Digital Video Microscopy for Colloidal Studies}.
\newblock \emph{Journal of Colloid and Interface Science}, 179\penalty0
  (1):\penalty0 298--310.
\newblock (\doi{10.1006/jcis.1996.0217})

\bibitem[Allan et~al.(2014)Allan, Caswell, Keim, Boulogne, Perry, and
  Uieda]{dan_allan_2014_12255}
Allan D, Caswell TA, Keim N, Boulogne F, Perry RW, Uieda L.
\newblock 2014, trackpy: Trackpy v0.2.4.
\newblock URL \url{https://doi.org/10.5281/zenodo.12255}

\bibitem[{Jafarzadeh} et~al.(2013){Jafarzadeh}, {Solanki}, {Feller}, {Lagg},
  {Pietarila}, {Danilovic}, {Riethm{\"u}ller}, and {Mart{\'\i}nez
  Pillet}]{2013A&A...549A.116J}
{Jafarzadeh} S, {Solanki} SK, {Feller} A, {Lagg} A, {Pietarila} A, {Danilovic}
  S, {Riethm{\"u}ller} TL, {Mart{\'\i}nez Pillet} V.
\newblock 2013, {Structure and dynamics of isolated internetwork Ca II H bright
  points observed by SUNRISE}.
\newblock \emph{\aap}, 549:\penalty0 A116.
\newblock (\doi{10.1051/0004-6361/201220089})

\bibitem[{Kianfar} et~al.(2018){Kianfar}, {Jafarzadeh}, {Mirtorabi}, and
  {Riethm{\"u}ller}]{2018SoPh..293..123K}
{Kianfar} S, {Jafarzadeh} S, {Mirtorabi} MT, {Riethm{\"u}ller} TL.
\newblock 2018, {Linear Polarization Features in the Quiet-Sun Photosphere:
  Structure and Dynamics}.
\newblock \emph{\solphys}, 293\penalty0 (8):\penalty0 123.
\newblock (\doi{10.1007/s11207-018-1341-2})

\bibitem[{Moreels} et~al.(2013){Moreels}, {Goossens}, and {Van
  Doorsselaere}]{2013A&A...555A..75M}
{Moreels} MG, {Goossens} M, {Van Doorsselaere} T.
\newblock 2013, {Cross-sectional area and intensity variations of sausage
  modes}.
\newblock \emph{\aap}, 555:\penalty0 A75.
\newblock (\doi{10.1051/0004-6361/201321545})

\bibitem[{Linnell Nemec} and {Nemec}(1985)]{1985AJ.....90.2317L}
{Linnell Nemec} AF, {Nemec} JM.
\newblock 1985, {A test of significance for periods derived using
  phase-dispersion-minimization techniques.}
\newblock \emph{\aj}, 90:\penalty0 2317--2320.
\newblock (\doi{10.1086/113936})

\bibitem[{O'Shea} et~al.(2001){O'Shea}, {Banerjee}, {Doyle}, {Fleck}, and
  {Murtagh}]{2001A&A...368.1095O}
{O'Shea} E, {Banerjee} D, {Doyle} JG, {Fleck} B, {Murtagh} F.
\newblock 2001, {Active region oscillations}.
\newblock \emph{\aap}, 368:\penalty0 1095--1107.
\newblock (\doi{10.1051/0004-6361:20010073})

\bibitem[{Torrence} and {Compo}(1998)]{1998BAMS...79...61T}
{Torrence} C, {Compo} GP.
\newblock 1998, {A Practical Guide to Wavelet Analysis.}
\newblock \emph{Bulletin of the American Meteorological Society}, 79\penalty0
  (1):\penalty0 61--78.
\newblock (\doi{10.1175/1520-0477(1998)079<0061:APGTWA>2.0.CO;2})

\bibitem[{Okamoto} and {De Pontieu}(2011)]{2011ApJ...736L..24O}
{Okamoto} TJ, {De Pontieu} B.
\newblock 2011, {Propagating Waves Along Spicules}.
\newblock \emph{\apjl}, 736\penalty0 (2):\penalty0 L24.
\newblock (\doi{10.1088/2041-8205/736/2/L24})

\bibitem[{Sekse} et~al.(2013){Sekse}, {Rouppe van der Voort}, and {De
  Pontieu}]{2013ApJ...764..164S}
{Sekse} DH, {Rouppe van der Voort} L, {De Pontieu} B.
\newblock 2013, {On the Temporal Evolution of the Disk Counterpart of Type II
  Spicules in the Quiet Sun}.
\newblock \emph{\apj}, 764\penalty0 (2):\penalty0 164.
\newblock (\doi{10.1088/0004-637X/764/2/164})

\bibitem[{Morton} et~al.(2014){Morton}, {Verth}, {Hillier}, and
  {Erd{\'e}lyi}]{2014ApJ...784...29M}
{Morton} RJ, {Verth} G, {Hillier} A, {Erd{\'e}lyi} R.
\newblock 2014, {The Generation and Damping of Propagating MHD Kink Waves in
  the Solar Atmosphere}.
\newblock \emph{\apj}, 784\penalty0 (1):\penalty0 29.
\newblock (\doi{10.1088/0004-637X/784/1/29})

\bibitem[{Morton} et~al.(2013){Morton}, {Verth}, {Fedun}, {Shelyag}, and
  {Erd{\'e}lyi}]{2013ApJ...768...17M}
{Morton} RJ, {Verth} G, {Fedun} V, {Shelyag} S, {Erd{\'e}lyi} R.
\newblock 2013, {Evidence for the Photospheric Excitation of Incompressible
  Chromospheric Waves}.
\newblock \emph{\apj}, 768\penalty0 (1):\penalty0 17.
\newblock (\doi{10.1088/0004-637X/768/1/17})

\bibitem[{He} et~al.(2009){He}, {Marsch}, {Tu}, and
  {Tian}]{2009ApJ...705L.217H}
{He} J, {Marsch} E, {Tu} C, {Tian} H.
\newblock 2009, {Excitation of Kink Waves Due to Small-Scale Magnetic
  Reconnection in the Chromosphere?}
\newblock \emph{\apjl}, 705\penalty0 (2):\penalty0 L217--L222.
\newblock (\doi{10.1088/0004-637X/705/2/L217})

\bibitem[{Jess} et~al.(2012){Jess}, {Pascoe}, {Christian}, {Mathioudakis},
  {Keys}, and {Keenan}]{2012ApJ...744L...5J}
{Jess} DB, {Pascoe} DJ, {Christian} DJ, {Mathioudakis} M, {Keys} PH, {Keenan}
  FP.
\newblock 2012, {The Origin of Type I Spicule Oscillations}.
\newblock \emph{\apjl}, 744\penalty0 (1):\penalty0 L5.
\newblock (\doi{10.1088/2041-8205/744/1/L5})

\bibitem[{Pereira} et~al.(2012){Pereira}, {De Pontieu}, and
  {Carlsson}]{2012ApJ...759...18P}
{Pereira} TMD, {De Pontieu} B, {Carlsson} M.
\newblock 2012, {Quantifying Spicules}.
\newblock \emph{\apj}, 759\penalty0 (1):\penalty0 18.
\newblock (\doi{10.1088/0004-637X/759/1/18})

\bibitem[{Kuridze} et~al.(2013){Kuridze}, {Verth}, {Mathioudakis},
  {Erd{\'e}lyi}, {Jess}, {Morton}, {Christian}, and
  {Keenan}]{2013ApJ...779...82K}
{Kuridze} D, {Verth} G, {Mathioudakis} M, {Erd{\'e}lyi} R, {Jess} DB, {Morton}
  RJ, {Christian} DJ, {Keenan} FP.
\newblock 2013, {Characteristics of Transverse Waves in Chromospheric Mottles}.
\newblock \emph{\apj}, 779\penalty0 (1):\penalty0 82.
\newblock (\doi{10.1088/0004-637X/779/1/82})

\bibitem[{Pietarila} et~al.(2011){Pietarila}, {Aznar Cuadrado}, {Hirzberger},
  and {Solanki}]{2011ApJ...739...92P}
{Pietarila} A, {Aznar Cuadrado} R, {Hirzberger} J, {Solanki} SK.
\newblock 2011, {Kink Waves in an Active Region Dynamic Fibril}.
\newblock \emph{\apj}, 739\penalty0 (2):\penalty0 92.
\newblock (\doi{10.1088/0004-637X/739/2/92})

\bibitem[{Morton} et~al.(2012){Morton}, {Verth}, {Jess}, {Kuridze}, {Ruderman},
  {Mathioudakis}, and {Erd{\'e}lyi}]{2012NatCo...3.1315M}
{Morton} RJ, {Verth} G, {Jess} DB, {Kuridze} D, {Ruderman} MS, {Mathioudakis}
  M, {Erd{\'e}lyi} R.
\newblock 2012, {Observations of ubiquitous compressive waves in the Sun's
  chromosphere}.
\newblock \emph{Nature Communications}, 3:\penalty0 1315.
\newblock (\doi{10.1038/ncomms2324})

\end{thebibliography}

\end{document}